\documentclass[12pt,a4paper]{article}
\usepackage{axodraw}
\usepackage{graphicx}
\usepackage{amsmath}
\usepackage{amsfonts}
\usepackage{amssymb}
\newcommand{\vs}{\vspace{-0.25cm}}

\newcommand{\fmd}{\,\mathrm{fm}^{-3}}

\begin{document}
%\hfill TUM/T39-02-??

\begin{center}

{\Large
\textbf{Chiral approach to nuclear matter: Role of two-pion\\ exchange with 
virtual delta-isobar excitation}\footnote{Work supported in part by BMBF and 
GSI.} }

\bigskip

{\large S. Fritsch, N. Kaiser, and W. Weise}\\

\bigskip

{\small Physik-Department, Technische Universit\"{a}t M\"{u}nchen,
D-85747 Garching, Germany\\
\smallskip

{\it email: nkaiser@physik.tu-muenchen.de}}

\end{center}

\bigskip

\begin{abstract}
We extend a recent three-loop calculation of nuclear matter in chiral 
perturbation theory by including the effects from two-pion exchange with 
single and double virtual $\Delta(1232)$-isobar excitation. Regularization 
dependent short-range contributions from pion-loops are encoded in a few 
NN-contact coupling constants. The empirical saturation point of 
isospin-symmetric nuclear matter, $\bar E_0 = -16\,$MeV, $\rho_0 =
0.16\,$fm$^{-3}$, can be well reproduced by adjusting the strength of a 
two-body term linear in density (and tuning an emerging three-body term
quadratic in density). The nuclear matter compressibility comes out as  $K
= 304\,$MeV. The real single-particle potential $U(p,k_{f0})$ is substantially 
improved by the inclusion of the chiral $\pi N\Delta$-dynamics: it
grows now monotonically with the nucleon momentum $p$. The effective nucleon
mass at the Fermi surface takes on a realistic value of $M^*(k_{f0})=0.88M$.
As a consequence of these features, the critical temperature of the liquid-gas
phase transition gets lowered to the value $T_c \simeq 15\,$MeV. In this work
we continue the complex-valued single-particle potential $U(p,k_f)+i\,
W(p,k_f)$ into the region above the Fermi surface $p>k_f$. The effects of
$2\pi$-exchange with virtual $\Delta$-excitation on the nuclear energy density
functional are also investigated. The effective nucleon mass associated with 
the kinetic energy density is  $\widetilde M^*(\rho_0)= 0.64M$. 
Furthermore, we find that the isospin properties of nuclear matter get
significantly improved by including the chiral $\pi N\Delta$-dynamics. Instead
of bending downward above $\rho_0$ as in previous calculations, the energy per
particle of pure neutron matter $\bar E_n(k_n)$ and the asymmetry
energy $A(k_f)$ now grow monotonically with density. In the density regime 
$\rho= 2\rho_n<0.2\,$fm$^{-3}$ relevant for conventional nuclear physics our
results agree well with sophisticated many-body calculations and 
(semi)-empirical values.   
\end{abstract}

\medskip

PACS: 12.38.Bx, 21.65.+f, 24.10.Cn, 31.15.Ew

\medskip
Keyword: Nuclear matter properties; Two-pion exchange with virtual
$\Delta(1232)$-isobar excitation

\section{Introduction and preparation}
In recent years a novel approach to the nuclear matter problem based on
effective field theory (in particular chiral perturbation theory) has
emerged. Its key element is a separation of long- and short-distance
dynamics and an ordering scheme in powers of small momenta. At nuclear matter
saturation density $\rho_0 \simeq 0.16\,$fm$^{-3}$ the Fermi momentum $k_{f0}$ 
and the pion mass $m_\pi$ are comparable scales ($k_{f0}\simeq 2 m_\pi$), and
therefore pions must be included as explicit degrees of freedom in the
description of the nuclear many-body dynamics. The contributions to the energy
per particle $\bar E(k_f)$ of isospin-symmetric nuclear matter as they 
originate from chiral pion-nucleon dynamics have been computed up to 
three-loop order in refs.\cite{lutz,nucmat}. Both calculations are able to 
reproduce correctly the empirical saturation point of nuclear matter by 
adjusting one single parameter (either a coupling $g_0+g_1 \simeq 3.23$
\cite{lutz} or a cut-off $\Lambda \simeq 0.65\,$GeV \cite{nucmat}) related to 
unresolved short-distance dynamics. The novel mechanism for saturation in
these approaches is a repulsive contribution to the energy per particle
generated by Pauli-blocking in second order (iterated) one-pion exchange. As
outlined in section 2.5 of ref.\cite{nucmat} this mechanism becomes
particularly transparent by taking the chiral limit $m_\pi = 0$. In that case
the interaction contributions to the energy per particle are completely
summarized by an attractive $k_f^3$-term and a repulsive $k_f^4$-term where
the parameter-free prediction for the coefficient of the latter is very close
to the one extracted from a realistic nuclear matter equation of state.  
  
The single-particle properties, represented by a complex-valued momentum and 
density dependent nucleon selfenergy $U(p,k_f)+i\,W(p,k_f)$, have been 
computed within our approach \cite{nucmat} in ref.\cite{pot}. The resulting 
potential depth $U(0,k_{f0}) = -53.2\,$MeV is in good agreement with that of 
the empirical nuclear shell \cite{bohr} or optical model \cite{hodgson}. 
However, the momentum dependence of the real single-particle potential 
$U(p,k_{f0})$ with its up- and downward bending (see Fig.\,3 in
ref.\cite{pot}) turns out to be too strong. As a consequence, the nominal 
value of the effective nucleon mass at the Fermi surface $p= k_{f0}$ would be
much too high: $M^*(k_{f0})\simeq  3M$. On the other hand, the single-particle 
properties around the Fermi surface are decisive for the spectrum of thermal
excitations and therefore they crucially influence the low temperature
behavior  of isospin-symmetric nuclear matter. The rather high critical
temperature $T_c \simeq 25.5\,$MeV for the liquid-gas phase transition
obtained in ref.\cite{liquidgas} is a visible manifestation of this intimate
relationship. 

While there is obviously room and need for improvement in our approach, one
must also note at the same time that the single-particle properties in the 
scheme of Lutz et al.\cite{lutz} (where explicit short-range terms are
iterated with pion-exchange) come out completely unrealistic 
\cite{lutzcontra}. The potential depth of $U(0,k_{f0}) = -20\,$MeV is by far 
too weakly attractive. Most seriously, the total single-particle energy 
$T_{\rm kin}(p)+ U(p,k_{f0})$ does not rise monotonically with the nucleon 
momentum $p$, thus implying a negative effective nucleon mass at the Fermi 
surface (see Fig.\,3 in ref.\cite{lutzcontra}). This ruins the behavior of 
nuclear matter at finite temperatures since a critical temperature of $T_c > 
40\,$MeV exceeds acceptable values by more than a factor of two.       

The isospin properties of nuclear matter  have also been investigated in 
ref.\cite{nucmat}. The prediction for the asymmetry energy at saturation 
density $A(k_{f0}) = 33.8\,$MeV is in good agreement with its empirical value. 
However, one finds a downward bending of $A(k_f)$ at densities $\rho > 
0.2\,$fm$^{-3}$. Such a behavior of the asymmetry energy $A(k_f)$
is presumably not realistic. The energy per particle of pure neutron 
matter $\bar E_n(k_n)$ as a function of the neutron density $\rho_n = 
k_n^3/3\pi^2$ shows a similar downward bending behavior (see Fig.\,8 in 
ref.\cite{nucmat}) and at lower neutron densities, there is only rough 
agreement with realistic neutron matter calculations. The mere fact that 
neutron matter came out to be unbound in ref.\cite{nucmat} with no further
adjusted parameter was however nontrivial. The isospin properties of nuclear 
matter in the scheme of Lutz et al. \cite{lutz} with a second free parameter 
adjusted, are qualitatively the same. The dashed curves in 
Figs.\,6,7 of ref.\cite{lutzcontra} display a downward bending of $\bar
E_n(k_n)$ and $A(k_f)$ at even lower densities, $\rho> 0.15\,$fm$^{-3}$. An 
extended version of that approach with pion-exchange plus two zero-range
NN-contact interactions iterated to second order and in total four adjustable
parameters has been  studied recently in ref.\cite{short}. The finding of that
work is that within such a complete fourth order calculation (thus exhausting
all possible terms up-to-and-including ${\cal O}(k_f^4)$) there is no optimal 
set of the four short-range parameters with which one could reproduce
simultaneously and accurately all semi-empirical properties of nuclear
matter. The conditions for a good neutron matter equation of state and equally
good single-particle properties (and consequently a realistic finite
temperature behavior) are in fact mutually exclusive in that approach. 

Calculations of nuclear matter based on the universal low-momentum 
nucleon-nucleon potential $V_{\rm low-k}$ have recently been performed in 
ref.\cite{kuckei}. The results obtained so far in Hartree-Fock or
Brueckner-Hartree-Fock approximation are unsatisfactory (see Figs.\,1,2 in
ref.\cite{kuckei}), since no saturation occurs in the equation of state. It
has been concluded that for the potential $V_{\rm low-k}$ the 
Brueckner-Hartree-Fock approximation is applicable only at very low
densities. These findings together with the identification of a successful 
saturation mechanism in the chiral approaches hint at the fact that the 
Brueckner ladder does not generate all relevant medium modifications which 
set in already at very low densities (typically at about one-tenth the
equilibrium density of nuclear matter, corresponding to Fermi momenta around 
$k_f \simeq m_\pi$).         
          
Up to this point the situation can be summarized as follows. Chiral two-pion
exchange restricted to nucleon intermediate states (basically the second-order
spin-spin and tensor force plus Pauli blocking effects), together with a 
single (finetuned) contact-term representing short-distance dynamics, is 
already surprisingly successful in binding and saturating nuclear matter and
reproducing key properties such as the asymmetry energy and the compression
modulus. However, the detailed behavior of the nucleon single-particle
potential and the density of states at the Fermi surface are not well
described at this order in the small-momentum expansion. 

Let us also comment on the relationship between our appoach to nuclear matter 
and the effective field theory treatment of (free) NN-scattering 
\cite{bira,entem,evgneu}. We are including  to the respective order the same 
long-range components from (one- and) two-pion exchange. The short-range 
NN-contact terms are however treated differently. Instead of introducing these
terms in a potential which is then iterated in a Lippmann-Schwinger equation, 
we adjust their strengths (in a perturbative calculation) to a few 
empirical nuclear matter properties. The resulting values of the coupling
constants are therefore not directly comparable with those of the
NN-potential. We note also that the important role of chiral two-pion exchange
for peripheral NN-scattering is well established \cite{entem,evgneu,nnpap}.

From the point of view of the driving pion-nucleon dynamics the previously
mentioned chiral calculations of nuclear matter \cite{lutz,nucmat,short} are
indeed still incomplete. They include only (S- and) P-wave Born terms but leave
out the excitation of the spin-isospin-3/2 $\Delta(1232)$-resonance, which is 
the prominent feature of low-energy $\pi N$-scattering. It is also well known 
that the two-pion exchange between nucleons with excitation of virtual 
$\Delta$-isobars generates most of the needed isoscalar central NN-attraction.
In phenomenological one-boson exchange models this part of the NN-interaction 
is often simulated  by a fictitious ''$\sigma$''-meson exchange. A 
parameter-free calculation of the isoscalar central potential $\widetilde 
V_C(r)$ generated by $2\pi$-exchange with single and double 
$\Delta$-excitation in ref.\cite{gerst} (see Fig.\,2 therein) agrees almost 
perfectly with the phenomenological ''$\sigma$''-exchange potential at 
distances $r>2\,$fm, but not at shorter distances. The more detailed behavior 
of the $2\pi$-exchange isoscalar central potential with single virtual
$\Delta$-excitation is reminiscent of the van-der-Waals potential. It has the
form \cite{gerst}: 
\begin{equation} \widetilde V_C^{(N\Delta)}(r) = - {3g_A^4 \over 64\pi^2
f_\pi^4 \Delta}\, {e^{-2x} \over r^6} (6+12x+10x^2+4x^3+x^4)\,, \nonumber 
\end{equation}
with $x=m_\pi r$ and the prefactor includes the spin-isospin (axial) 
polarizability of the nucleon \cite{ericson}, $g_A^2/ f_\pi^2\Delta=5.2\,
$fm$^3$, from the  virtual $N \to \Delta(1232) \to N$ transition. The familiar
$r^{-6}$-dependence of the non-relativistic van-der-Waals interaction emerges
in the chiral limit, $m_\pi = 0$.   

A consideration of mass scales also suggests to include the $\Delta(1232)
$-isobar as an explicit degree of freedom in nuclear matter calculations. The 
delta-nucleon mass splitting of $\Delta = 293\,$MeV is comparable to 
the Fermi momentum $k_{f0} \simeq 262$\,MeV at nuclear matter 
saturation density. Propagation effects of virtual $\Delta(1232)$-isobars can 
therefore be resolved at the densities of interest. Based on these scale 
arguments we adopt a calculational scheme in which we count the Fermi momentum
$k_f$, the pion mass $m_\pi$  and the $\Delta N$-mass splitting $\Delta$ 
simultaneously as ''small scales''. The non-relativistic treatment of the 
nuclear matter many-body problem naturally goes conform with such an expansion 
in powers of small momenta. Relativistic corrections are relegated to 
higher orders in this expansion scheme. The leading contributions from 
$2\pi$-exchange with virtual $\Delta$-excitation to the energy per particle
(or the single-particle potential) are generically of fifth power in the small
momenta  ($k_f,m_\pi,\Delta$). With respect to the counting in small momenta
the effects from irreducible $2\pi$-exchange evaluated in 
ref.\cite{nucmat,pot,liquidgas} belong to the same order. However, since the
$\pi N\Delta$-coupling constant is about twice as large as the $\pi
NN$-coupling constant one can expect that the $\Delta$-driven $2\pi$-exchange
effects are the dominant ones. The importance of $\Delta(1232)$-degrees of
freedom has also been pointed out in the ''ab-initio'' calculations of the 
Illinois group \cite{akmal,pieper}.   

The purpose of the present paper is to present a calculation of
(isospin-symmetric and  isospin-asymmetric) nuclear matter which includes
systematically all effects from $2\pi$-exchange with virtual 
$\Delta$-excitation up to three-loop order in the energy density. The 
contributions to the energy per particle (or the single-particle potential) 
can be classified as two-body terms and three-body terms. Two-body terms can 
be directly expressed through the NN-scattering T-matrix (i.e. the 
NN-potential in momentum space). Three-body terms on the other hand can be 
interpreted as Pauli-blocking effects on the two-body terms imposed by the 
filled Fermi-sea of nucleons. The notion of ''three-body term'' is taken here
in a more general context, namely in the sense that three nucleons in the 
Fermi sea participate in interactions. The NN T-matrix generated by the 
in general ultra-violet divergent pion-loop diagrams requires regularization 
(and renormalization). We adopt here a suitably subtracted dispersion-relation 
representation of the T-matrix where this procedure is accounted for by a few 
subtraction constants. The latter constants are understood to encode 
unresolved short-distance NN-dynamics. The associated $k_f^3$- and 
$k_f^5$-terms in the energy per particle are then adjusted to some empirical 
property of nuclear matter (e.g. the maximal binding energy of $16\,$MeV). 

Our paper is organized as follows: In section 2 we start with the equation of 
state of isospin-symmetric nuclear matter. We evaluate analytically the 
three-loop in-medium diagrams generated by the chiral $\pi N\Delta$-dynamics
and perform the necessary adjustment of short-range parameters. Section 3 
deals with the real single-particle potential $U(p,k_f)$ whose improved 
momentum dependence turns out to be a true prediction. In section 4 we 
reconsider the imaginary single-particle potential $W(p,k_f)$ of
ref.\cite{pot}, but now extended into the region above the Fermi surface 
$p>k_f$. Section 5 is devoted to the effective nucleon mass $\widetilde
M^*(\rho)$ and the strength function of the $(\vec \nabla \rho)^2$-term in the
nuclear energy density functional. In section 6 we extend our calculation of
isospin-symmetric nuclear matter to finite temperatures $T$. The main interest
lies there on the critical temperature $T_c$ of the first-order liquid-gas
phase transition for which we find an improved value of $T_c \simeq 15\,$MeV. 
Sections 7, 8 and 9 deal with the equation of state of pure neutron matter 
$\bar E_n(k_n)$, the asymmetry energy $A(k_f)$ and the isovector
single-particle potential $U_I(p,k_f)$. These three quantities reveal the
isospin properties of the underlying $\pi N \Delta$-dynamics. Explicit
inclusion of the $\Delta(1232)$-degree of freedom leads to a substantial 
improvement: the notorious downward bending of $\bar E_n(k_n)$ and $A(k_f)$ 
observed in previous chiral calculations is now eliminated. Finally, section 10
ends with a summary and some concluding remarks.           
  
\section{Equation of state of isospin-symmetric nuclear matter}
We start the discussion with the equation of state of isospin-symmetric
nuclear matter for which one has a fairly good knowledge of the empirical
saturation point. We first write down the contributions to the energy per
particle $\bar E(k_f)$ as they arise from $2\pi$-exchange with single and
double virtual $\Delta$-isobar excitation.

\begin{figure}
\begin{center}
\includegraphics[scale=1.,clip]{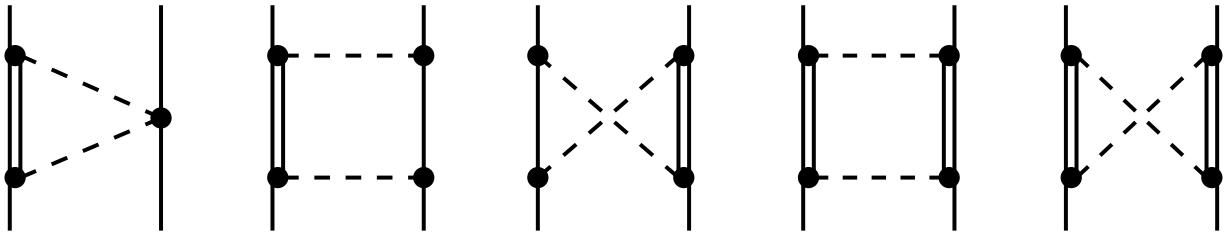}
\end{center}\vspace{-0.4cm}
\caption{One-loop two-pion exchange diagrams with single and double
$\Delta(1232)$-isobar excitation. Diagrams for which the role of both nucleons
is interchanged are not shown.}
\end{figure}

Fig.\,1 shows the relevant one-loop triangle, box and crossed box diagrams
contributing to the NN T-matrix (or the momentum space potential). The finite
parts of these diagrams have been evaluated analytically in section 3 of
ref.\cite{gerst} employing the usual non-relativistic $\Delta \leftrightarrow 
\pi N$ transition vertices and $\Delta$-propagator (see eq.(4) in 
ref.\cite{gerst}). By closing the two open nucleon lines to either two rings or
one ring one gets (in diagrammatic representation) the Hartree or Fock 
contribution to the energy density of nuclear matter. The Hartree contribution 
to the energy per particle evidently goes linear with the nucleon density 
$\rho=  2k_f^3/3\pi^2$, namely $\bar E(k_f)^{(2H)} = -V_C(0)\, \rho/2$ with 
$V_C(0)$ the isoscalar central NN-amplitude at zero momentum transfer
\cite{gerst}. The Fock contribution on the other hand is obtained by
integrating the spin- and isospin-contracted T-matrix (depending on the
momentum transfer variable $|\vec p_1 - \vec p_2|$) over the product of two
Fermi spheres $|\vec p_{1,2}|<k_f$ of radius $k_f$. We separate regularization
dependent short-range parts  in the T-matrix (originating from the divergences
of the loop diagrams) from the unique long-range terms with the help of a
twice-subtracted dispersion relation. The resulting subtraction constants give
rise to a contribution to the energy per particle of the form:
\begin{equation} \bar E(k_f)^{(ct)}= B_3 {k_f^3 \over M^2} + B_5 {k_f^5 \over
M^4}\,,  \end{equation}
where $B_3$ and $B_5$ are chosen for convenience as dimensionless. 
$M=939\,$MeV stands for the (average) nucleon mass. Note that 
eq.(1) is completely equivalent to the contribution of a momentum independent
and $p^2$-dependent NN-contact interaction. We interpret the parameters
$B_{3,5}$ to subsume all unresolved short-distance NN-dynamics relevant for
isospin-symmetric nuclear matter at low and moderate densities. The long-range
parts of the $2\pi$-exchange (two-body) Fock diagrams can be expressed as:    
\begin{eqnarray} \bar E(k_f)^{(2F)}&=& {1 \over 8\pi^3} \int_{2m_\pi}^{\infty}
\!\! d\mu\,{\rm Im}(V_C+3W_C+2\mu^2V_T+6\mu^2W_T) \bigg\{ 3\mu k_f -{4k_f^3
\over 3\mu }\nonumber \\ &&+{8k_f^5 \over 5\mu^3 } -{\mu^3 \over 2k_f}-4\mu^2
\arctan{2k_f\over\mu} +{\mu^3 \over 8k_f^3}(12k_f^2+\mu^2) \ln\bigg( 1+{4k_f^2
\over \mu^2} \bigg) \bigg\} \,, \end{eqnarray} 
where Im$V_C$, Im$W_C$, Im$V_T$ and Im$W_T$ are the spectral functions of the
isoscalar and isovector central and tensor NN-amplitudes, respectively. 
Explicit expressions of these imaginary parts for the contributions of the 
triangle diagram with single $\Delta$-excitation and the box diagrams with 
single and double $\Delta$-excitation can be easily constructed from the 
analytical formulas given in section 3 of ref.\cite{gerst}. The $\mu$- and 
$k_f$-dependent weighting function in eq.(2) takes care that at low and
moderate densities this spectral integral is dominated by low invariant
$\pi\pi$-masses $2m_\pi< \mu <1\,$GeV. The contributions to the energy per
particle from irreducible $2\pi$-exchange (with only nucleon intermediate
states) can also be cast into the form eq.(2). The corresponding
non-vanishing spectral functions read \cite{nnpap}: 
\begin{equation} {\rm Im}W_C(i\mu) = {\sqrt{\mu^2-4m_\pi^2} \over 3\pi 
\mu (4f_\pi)^4} \bigg[ 4m_\pi^2(1+4g_A^2-5g_A^4) +\mu^2(23g_A^4-10g_A^2-1) + 
{48 g_A^4 m_\pi^4 \over \mu^2-4m_\pi^2} \bigg] \,, \end{equation}
\begin{equation} {\rm Im}V_T(i\mu) = - {6 g_A^4 \sqrt{\mu^2-4m_\pi^2} \over 
\pi  \mu (4f_\pi)^4}\,. \end{equation}
The dispersion integrals $\int_{2 m_\pi}^\infty d\mu\,{\rm Im}(\dots)$ in this
and all following sections are understood to include the contributions from
irreducible $2\pi$-exchange (with only nucleon intermediate states).  

\begin{figure}
\begin{center}
\includegraphics[scale=1.0,clip]{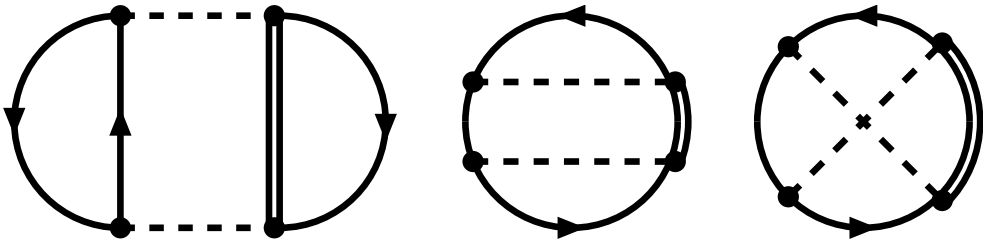}
\end{center}\vspace{-0.2cm}
\caption{Hartree and Fock three-body diagrams related to $2\pi$-exchange with 
single virtual $\Delta$-isobar excitation. They represent interactions between 
three nucleons in the Fermi sea. In the case of isospin-symmetric nuclear 
matter the isospin factors of these diagrams are 8, 0, and 8, in the order 
shown. The combinatoric factor is 1 for each diagram.} 
\end{figure}
Next, we come to the three-body terms which arise from Pauli blocking of
intermediate nucleon states (i.e., from the $-2\pi \theta(k_f -|\vec p\,|)$
terms of the in-medium nucleon propagators \cite{nucmat}). The corresponding 
closed Hartree and Fock diagrams with single virtual $\Delta$-excitation are
shown in Fig.\,2. In the case of isospin-symmetric nuclear matter their
isospin factors are 8, 0, and 8, in the order shown. For the three-loop
Hartree diagram the occurring integral over the product of three Fermi spheres
of radius $k_f$ can be solved in closed form and the contribution to the
energy per particle reads: 
\begin{equation} \bar E(k_f)^{(3H)}={g_A^4 m_\pi^6 \over \Delta(2\pi f_\pi)^4}
\bigg[ {2\over3}u^6(1+\zeta) +u^2-3u^4+5u^3 \arctan2u-{1\over 4}(1+9u^2)
\ln(1+4u^2) \bigg] \,, \end{equation}  
with the abbreviation $u=k_f/m_\pi$ where $m_\pi=135\,$MeV stands for the 
(neutral) pion mass. The delta-propagator shows up in this expression merely 
via the (reciprocal) mass-splitting $\Delta = 293\,$MeV.
Additional corrections to the delta-propagator coming from differences of
nucleon kinetic energies etc. will make a contribution at least one order
higher in the small-momentum expansion. In eq.(5) we have already inserted the 
empirically well-satisfied relation $g_{\pi N \Delta} = 3g_{\pi N}/\sqrt{2}$
for the $\pi N \Delta$-coupling constant together with the Goldberger-Treiman
relation $g_{\pi N}= g_A M/f_\pi$ (see e.g. eq.(5) in ref.\cite{gerst} for the
$\Delta \to N \pi$ decay width). As usual $f_\pi = 92.4\,$MeV denotes the 
weak pion decay constant and we choose the value $g_A=1.3$ in order to have a
pion-nucleon coupling constant of $g_{\pi N} = 13.2$ \cite{pavan}. Via the 
parameter $\zeta$ we have already included in eq.(5) the contribution of an 
additional three-nucleon contact interaction $\sim (\zeta g_A^4/\Delta f_\pi^4
)\,(\bar NN)^3$. One notices that the $2\pi$-exchange three-body Hartree
diagram shows in the chiral limit $m_\pi = 0$ the same quadratic dependence
on the nucleon density. In that limit the momentum dependent $\pi N
\Delta$-interaction vertices get canceled by the pion-propagators and thus one
is effectively dealing with a zero-range three-nucleon contact-interaction. It
is important to point out that this equivalence holds only after taking the
spin-traces but not at the level of the (spin- and momentum dependent)
$2\pi$-exchange  three-nucleon interaction. The contribution of the (right)
three-body Fock diagram in Fig.\,2 to the energy per particle reads: 
\begin{equation} \bar E(k_f)^{(3F)}=-{3g_A^4 m_\pi^6 u^{-3}\over 4\Delta(4\pi
f_\pi)^4 }\int_0^u \!\! dx \Big[ 2G^2_S(x,u)+G^2_T(x,u)\Big] \,, \end{equation}
where we have introduced the two auxiliary functions:
\begin{eqnarray} G_S(x,u) &=& {4ux \over 3}( 2u^2-3) +4x\Big[
\arctan(u+x)+\arctan(u-x)\Big] \nonumber \\ && + (x^2-u^2-1) \ln{1+(u+x)^2
\over  1+(u-x)^2} \,,\end{eqnarray}
\begin{eqnarray} G_T(x,u) &=& {ux\over 6}(8u^2+3x^2)-{u\over
2x} (1+u^2)^2  \nonumber \\ && + {1\over 8} \bigg[ {(1+u^2)^3 \over x^2} -x^4 
+(1-3u^2)(1+u^2-x^2)\bigg] \ln{1+(u+x)^2\over  1+(u-x)^2} \,.\end{eqnarray}
Evidently, the three-body Fock term in eq.(6) is attractive.  The remaining
contributions to $\bar E(k_f)$ from the (relativistically improved) kinetic
energy, from the $1\pi$-exchange Fock diagram and from the iterated
$1\pi$-exchange Hartree and Fock diagrams have been written down in eqs.(5-11) 
of ref.\cite{nucmat}. The strongly attractive contribution from iterated 
$1\pi$-exchange linear in the density and the cutoff $\Lambda$ (see eq.(15) in
ref.\cite{nucmat}) is now of course not counted extra since $B_3$ in eq.(1)
collects all such possible terms. Adding all pieces we arrive at the full
energy per particle $\bar E(k_f)$ at three-loop order. It involves three
parameters, $B_3$ and $B_5$ of the two-body contact term eq.(1) and $\zeta$ 
which controls a three-body contact term in eq.(5).    

Let us first look at generic properties of the nuclear matter equation of 
state in our calculation. Binding and saturation occurs in a wide range of
the two adjustable parameters $B_{3,5}$. However, with the full strength 
($\zeta=0$) of the repulsive $\rho^2$-term from the $2\pi$-exchange three-body
Hartree diagram (see eq.(5)) the saturation curve rises much too steeply with 
increasing density. This causes a too low saturation density $\rho_0$ and
a too high nuclear matter compressibility, $K> 350\,$MeV. We cure this problem
in a minimal way by introducing an attractive three-body contact term. With 
$\zeta =-3/4$ the remaining repulsive $\rho^2$-term in eq.(5) gets canceled by
an analogous attractive contribution from the three-body Fock diagram. 
Clearly, the need for introducing an attractive three-body contact term into 
our calculation points to some short-distance physics whoose dynamical origin 
lies outside the present framework of perturbative chiral pion-nucleon 
interactions. It will become clear in the following sections that the 
predictive power of our calculation is nevertheless not reduced by this 
procedure. 

\begin{figure}
\begin{center}
\includegraphics[scale=0.55,clip]{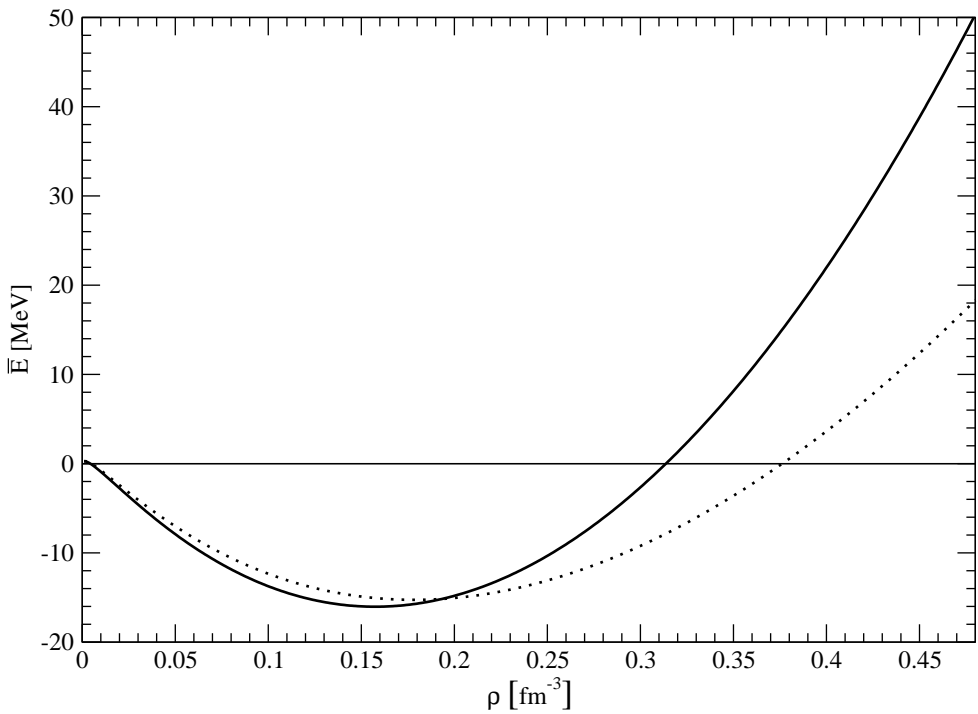}
\end{center}\vspace{-0.2cm}
\caption{The energy per particle $\bar E(k_f)$ of isospin-symmetric nuclear
matter as a function of the nucleon density $\rho= 2k_f^3/3\pi^2$. The dashed
line refers to the result of ref.\cite{nucmat}, with only pions and nucleons as
active degrees of freedom. The full line includes effects from $2\pi$-exchange 
with virtual $\Delta$-excitation. The short-range parameters are $B_3=-7.99$
and $B_5=0$.}    
\end{figure}

We fix the minimum of the saturation curve $\bar E(k_f)$ to the
value $\bar E_0= -16.0\,$MeV. With $B_3$ adjusted to the value $B_3 = -7.99$
and $B_5$ taken to be zero, $B_5 = 0$, the full line in Fig.\,3 results. The 
predicted value of the saturation density is $\rho_0= 0.157\,$fm$^{-3}$,
corresponding to a Fermi momentum of $k_{f0} =261.6\,$MeV $=1.326\,$fm$^{-1}$.
This is very close to the semi-empirical value $\rho_0=0.158\fmd$ obtained by
extrapolation from inelastic electron scattering off heavy nuclei \cite{sick}.
The decomposition of the negative binding energy $\bar E_0 = -16.0\,$MeV into 
contributions of second, third, fourth and fifth power in small momenta reads:
$\bar E_0 = (21.9-145.5+107.8-0.2)\,$MeV with the typical balance between 
large third and fourth order terms \cite{nucmat}. The very small fifth order
term splits furthermore as $(-13.8 +13.6)\,$MeV into the contribution from the
three-body contact-interaction (proportional to $\zeta = -3/4$) and a
remainder. Evidently, since $\bar E_0 = -16.0\,$MeV is a small number that
needs to be finetuned in our calculation there remains the question of the 
''convergence'' of the small momentum expansion.   
The nuclear matter compressibility $K = k_{f0}^2 \bar E''(k_{f0})$ related to
the curvature of the saturation  curve at its minimum comes out as
$K=304\,$MeV. This number is somewhat high but still acceptable, since it
exceeds e.g. the value $K=272\,$MeV obtained in the relativistic mean-field
model of ref.\cite{vretenar} only by $12\%$. Inspection of Fig.\,3 shows that
the saturation curve is well approximated by a shifted parabola $\bar
E(k_f) = \bar E_0 (2\rho_0-\rho) \rho/\rho_0^2$ with a second zero-crossing at
twice nuclear matter density $2\rho_0$. This leads to a compressibility
estimate of $K= -18 \bar E_0= 288\,$MeV, not far from the calculated value. We
have also studied variations of the parameter $B_5$. Within the limits set by
a stable saturation point (and realistic single-particle properties, see next
section), the effects on the compressibility $K$ are marginal 
(less than 10\,MeV reduction). Therefore we stay with the minimal choice
$B_5=0$ (together with $B_3 = -7.99$ and $\zeta=-3/4$). The dashed line in 
Fig.\,3 shows for comparison the equation of state resulting from our previous 
chiral calculation \cite{nucmat} with no $\pi N \Delta$-dynamics included. In 
that work the saturation density $\rho_0= 0.178\,$fm$^{-3}$ came out somewhat
too high, but the compressibility $K=255\,$MeV had a better value. The
stronger rise of the full curve in Fig.\,3 with density $\rho$ is a
consequence of including higher order terms in the (small-momentum)
$k_f$-expansion.  
\section{Real part of single-particle potential}
In this section we discuss the real part $U(p,k_f)$ of the single-particle
potential. As outlined in ref.\cite{pot} the contributions to the (real)
nuclear mean-field  $U(p,k_f)$ can be classified as two-body and three-body
potentials. The parameters $B_{3,5}$ introduced in eq.(1) reappear in a 
contribution to the two-body potential of the form:   
\begin{equation} U_2(p,k_f)^{(ct)}= 2B_3 {k_f^3 \over M^2} + B_5 {k_f^3 \over 
3M^4}  (3k_f^2+5p^2)\,.  \end{equation}
Its density- and momentum-dependence is completely fixed by the
Hugenholtz-Van-Hove theorem \cite{vanhove} and a sum rule which connects it to
the energy per particle $\bar E(k_f)^{(ct)}$ (see eqs.(5,7) in
ref.\cite{pot}). The Fock diagrams of $2\pi$-exchange with virtual
$\Delta$-excitation give rise to a contribution to the two-body potential
which can be written as a (subtracted) dispersion integral:   
\begin{eqnarray} U_2(p,k_f)^{(F)}&=& {1 \over 2\pi^3} \int_{2m_\pi}^{\infty}
\!\!d\mu \,{\rm Im}(V_C+3W_C+2\mu^2V_T+6\mu^2W_T)\nonumber \\ && \times
\bigg\{ \mu k_f +{2k_f^3\over 15\mu^3} (3k_f^2+5p^2)-{2k_f^3 \over 3\mu}-\mu^2
\arctan{k_f+ p\over\mu} \nonumber \\ && -\mu^2 \arctan{k_f-p\over \mu}+ { \mu
\over 4p}(\mu^2+k_f^2-p^2)\ln{\mu^2+ (k_f+p)^2 \over \mu^2+(k_f-p)^2} \bigg\} 
\,. \end{eqnarray}
By opening a nucleon line in the three-body diagrams of Fig.\,2 one gets (per
diagram) three different contributions to the three-body potential. In the 
case of the (left) Hartree diagram they read altogether:   
\begin{eqnarray} U_3(p,k_f)^{(H)}&=&{g_A^4 m_\pi^6 \over \Delta(2\pi f_\pi)^4} 
\bigg\{2u^6(1+\zeta)+ u^2-7u^4-{1\over 4}(1+9u^2)\ln(1+4u^2) \nonumber \\ &&
+5u^3 \Big[\arctan2u+\arctan(u+x)+ \arctan(u-x)\Big] 
\nonumber \\ && +{u^3\over 2x}(2x^2-2u^2-3)
\ln{1+(u+x)^2 \over  1+(u-x)^2 } \bigg\} \,, \end{eqnarray} 
with the abbreviation $x = p/m_\pi$. Note that the paramter $\zeta = -3/4$
related to the three-body contact-interaction has no influence on the momentum
dependence of the single-particle potential. On the other hand the (right) 
Fock diagram in Fig.\,2 generates a total contribution to the three-body
potential of the form:   
\begin{eqnarray} U_3(p,k_f)^{(F)}&=&-{g_A^4m_\pi^6 x^{-2}\over 4\Delta(4\pi
f_\pi)^4} \bigg\{ 2G_S^2(x,u)+G_T^2(x,u)\nonumber \\ && +\int_0^u\!\!d\xi
\bigg[ 4G_S(\xi,u) {\partial G_S(\xi,x) \over \partial x} +2G_T(\xi,u) 
{\partial G_T(\xi,x) \over \partial x} \bigg] \bigg\}\,, \end{eqnarray}
with $G_{S,T}(x,u)$ defined in eqs.(7,8). The real single-particle potential
$U(p,k_f)$ is completed by adding to the terms eqs.(9-12) the contributions
from $1\pi$-exchange and iterated $1\pi$-exchange written down in eqs.(8-13) of
ref.\cite{pot}. The slope of the real single-particle potential $U(p,k_f)$ at 
the Fermi surface $p=k_f$ determines the effective nucleon mass (in the 
nomenclature of ref.\cite{mahaux}, the product of ''$k$-mass" and ''$E$-mass" 
divided by the free nucleon mass $M=939\,$ MeV) via a relation:
\begin{equation} M^*(k_f)  = M \bigg[ 1- {k_f^2 \over 2M^2} + {M\over k_f}\,
{\partial U(p,k_f) \over \partial p}\bigg|_{p=k_f}\bigg]^{-1} \,.\end{equation}
The second term $-k_f^2/2M^2$ in the square brackets stems from the
relativistic correction $-p^4/8M^3$ to the kinetic energy.  

\begin{figure}
\begin{center}
\includegraphics[scale=0.56,clip]{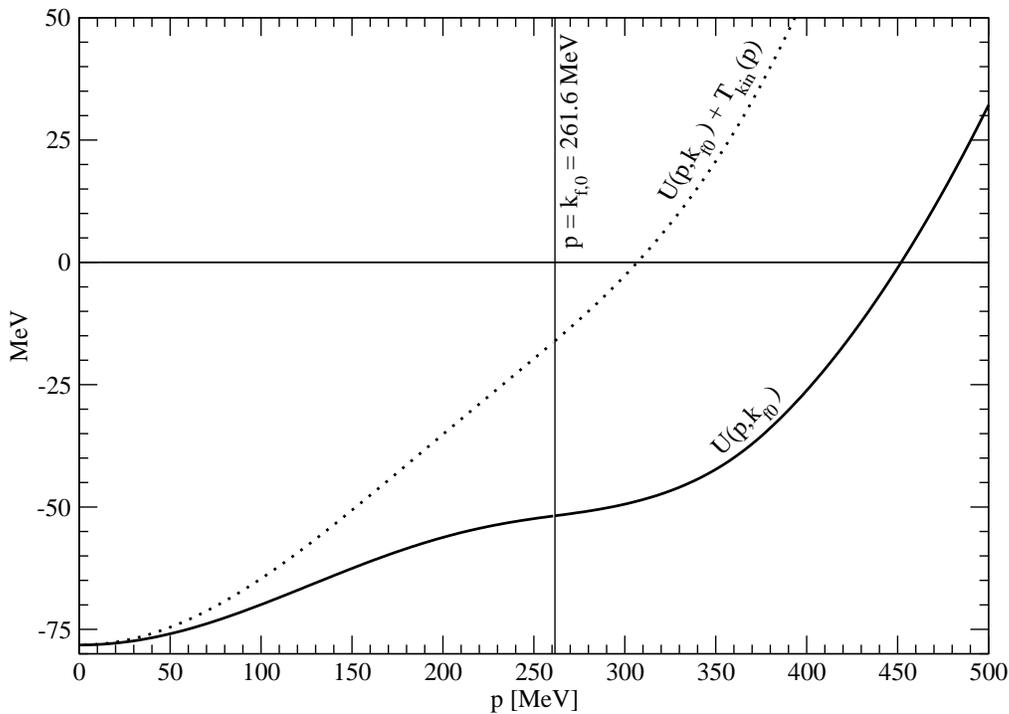}
\end{center}\vspace{-0.2cm}
\caption{Full line: real part of the single-particle potential 
$U(p,k_{f0})$ at saturation density $k_{f0}= 261.6\,$MeV as a function of the
nucleon momentum $p$. The dotted line includes in addition the
relativistically improved kinetic energy $T_{\rm kin}(p) = p^2/2M-p^4/8M^3$. 
Both curves are extended into the region above the Fermi surface $p> k_{f0}$.}
\end{figure}

The full line in Fig.\,4 shows the real part of the single-particle potential 
$U(p,k_{f0})$ at saturation density $k_{f0}= 261.6\,$MeV as a function of the
nucleon momentum $p$. The dotted line includes in addition the 
relativistically improved kinetic energy $T_{\rm kin}(p) = p^2/2M-p^4/8M^3$. 
With the parameters $B_5=0$, $\zeta=-3/4$ fixed and $B_3 = -7.99$ adjusted in 
section 1 to the binding energy at equilibrium we find a potential depth of
$U(0,k_{f0})= -78.2\,$MeV. This is very close to the result $U(0,k_{f0})\simeq
-80\,$MeV of the relativistic Dirac-Brueckner approach of ref.\cite{rolf}. For
comparison, the calculation of ref.\cite{grange} based on the phenomenological
Paris NN-potential finds a somewhat shallower potential depth of $U(0,k_{f0})
\simeq -64\,$MeV. One observes that with the chiral $\pi N\Delta$-dynamics
included, the real single-particle potential $U(p,k_{f0})$ grows
monotonically with the nucleon momentum $p$. The downward bending above $p=
180\,$MeV displayed in Fig.\,3 of ref.\cite{pot} is now eliminated. The slope
at the Fermi surface $p=k_{f0}$ translates into an effective nucleon mass of
$M^*(k_{f0}) = 0.88M$. This is now a realistic value compared to $M^*(k_{f0})
\simeq 3M$ obtained in our previous calculation \cite{pot,liquidgas} 
without any explicit $\Delta$-isobars. Note also that the chiral approach of
ref.\cite{short} (where both explicit short-range terms and pion-exchange are
iterated) has found the lower bound $M^*(k_{f0}) >1.4M$.  

The dotted curve in Fig.\,4 for the total single-particle energy $T_{\rm 
kin}(p) + U(p,k_{f0})$ hits the value $\bar E(k_{f0})=\bar E_0=-16\,$MeV at 
the Fermi surface $p= k_{f0}$, as required by the Hugenholtz-Van-Hove theorem
\cite{vanhove}. This  important theorem holds strictly in our (perturbative)
calculation, whereas  (non-perturbative) Brueckner-Hartree-Fock approaches
often fail to respect it \cite{bozek}. In Fig.\,4 we have also extended both 
curves into the region above the Fermi surface $p > k_{f0}$. In general this
extension is not just an analytical continuation of the potential from below 
the Fermi surface. Whereas eqs.(9-12) for the contributions from
$2\pi$-exchange with $\Delta$-excitation apply in both regions, there are
non-trivial changes in the expressions from iterated $1\pi$-exchange. These
modifications are summarized in the Appendix. The smooth rise of $U(p,k_{f0})$
as it crosses the Fermi surface and proceeds up to $p \simeq 400\,$MeV is 
compatible with other calculations \cite{rolf,grange}. Beyond this momentum
scale one presumably exceeds the limits of validity of the present chiral 
perturbation theory calculation of nuclear matter. 

\begin{figure}
\begin{center}
\includegraphics[scale=0.55,clip]{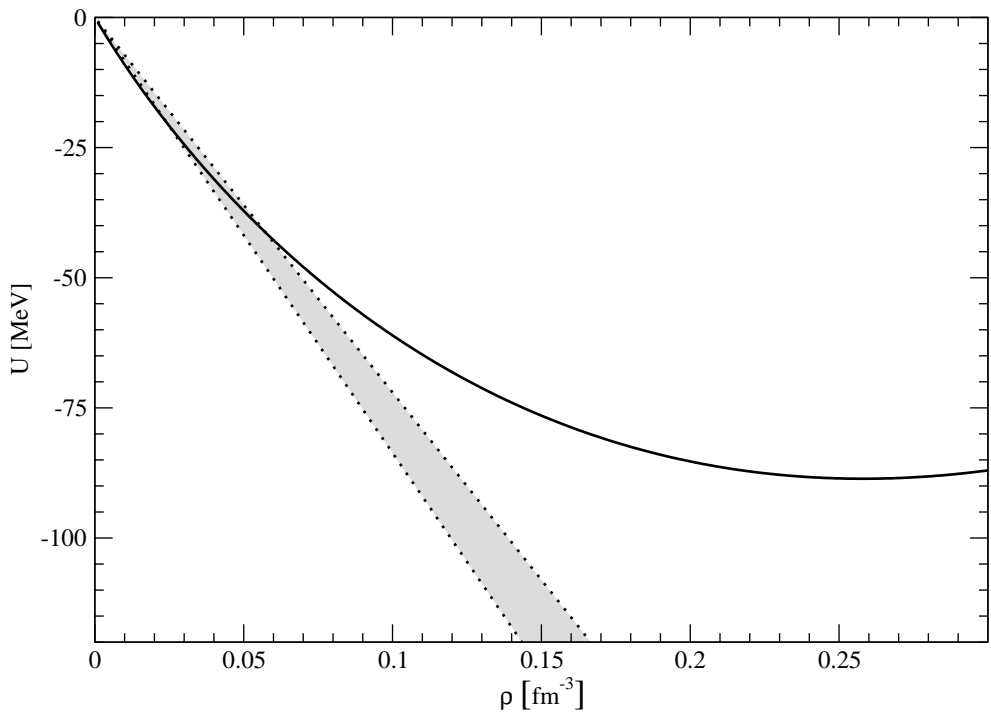}
\end{center}\vspace{-0.2cm}
\caption{The full line shows the real part of the single-particle potential
$U(0,k_f)$ at nucleon momentum $p=0$ versus the density $\rho= 2k_f^3/3\pi^2$.
The band is obtained from the universal low-momentum NN-potential
$V_{\rm low-k}$ in linear density approximation.} 
\end{figure}

The full line in Fig.\,5 shows the potential depth $U(0,k_f)$ for a nucleon 
at the bottom of the Fermi sea as a function of the nucleon density $\rho =
2k_f^3/3\pi^2$. The band spanned by the dotted lines stems from the universal 
low-momentum NN-potential $V_{\rm low-k}$ \cite{vlowk} in linear density
approximation. In this approximation the potential depth simply reads:
\begin{equation} U(0,k_f) = {3 \pi \rho \over 2M} \Big[ V_{\rm low-k}^{
(^1\!S_0)}(0,0)+ V_{\rm low-k}^{(^3\!S_1)}(0,0) \Big] \,, \end{equation}
with $V_{\rm low-k}^{(^1\!S_0)}(0,0)\simeq -1.9\,$fm and $V_{\rm low-k}^{(^3\!
S_1)}(0,0) \simeq -(2.2\pm 0.3)\,$fm \cite{vlowk,gerry,feldmeier}, the two 
S-wave potentials at zero momentum. It is interesting to observe that both
potential depths agree fairly well at low densities, $\rho \leq 
0.07\,$fm$^{-3}$. This agreement is by no means trivial since $V_{\rm low-k}$
is constructed to reproduce accurately the low-energy NN-scattering data
(phase-shifts and mixing angles) whereas our adjustment of the $B_3$-term
(linear in density) is made at saturation density $\rho_0=0.157\, $fm$^{-3}$. 
It is evident from Fig.\,5 that a linear extrapolation does not work from zero
density up to nuclear matter saturation density. Strong curvature effects set
in already at Fermi momenta around $k_f\simeq m_\pi$ if (and only if)
pion-dynamics is treated explicitly.\footnote{As example for the extreme
inherent non-linearities, consider the formula for the three-body potential
$U_3(0,k_f)^{(H)}$ in eq.(11). Its mathematical Taylor-series expansion
converges only for $k_f <m_\pi/2$. This corresponds to tiny densities, $\rho <
0.0027\,$fm$^{-3}$.} We note also that an ''improved'' determination of the 
potential depth $U(0,k_f)$ from $V_{\rm low-k}$ where one takes into account
its momentum dependence in the repulsive Fock contribution leads to concave 
curves which bend below the straight dotted lines in Fig.\,5. In the case
$V_{\rm low-k}^{(^3\!S_1)}(0,0)=-1.9\,$fm the potential depth $U(0,k_{f0})$ at 
saturation density would increase to $-132.6\,$MeV (compared to $-113.4\,$MeV 
in linear density approximation). The present observations concerning the 
potential depth $U(0,k_f)$ may indicate why calculations based on $V_{\rm 
low-k}$ did so far not find saturation of nuclear matter \cite{kuckei}. It 
seems that the Brueckner ladder does not generate all relevant medium 
modifications which set in already at rather low densities  $k_f \simeq m_\pi$ 
(if the pion-dynamics is treated explicitly).   

\section{Imaginary part of single-particle potential}
In this section, we reconsider the imaginary part $W(p,k_f)$ of the
single-particle potential. To the three-loop order we are working here it is 
still given completely by iterated $1\pi$-exchange with no contribution from 
the $\pi N \Delta$-dynamics. The new aspect here is the extension into the
region above the Fermi surface $p> k_f$, which is not an analytical
continuation from below the Fermi surface. As outlined in ref.\cite{pot} the
contributions to $W(p,k_f)$ can be classified as two-body, three-body and
four-body terms. From the Hartree diagram of iterated $1\pi$-exchange one
finds altogether (in the region above the Fermi surface  $p> k_f$):   
\begin{eqnarray} W(p,k_f)^{(H)} & =& {\pi g_A^4M m_\pi^4 \over (4\pi f_\pi)^4}
\Bigg\{ \Big( 9+6u^2+{4u^3 \over x}-2x^2 \Big) \ln[1+(u+x)^2]
\nonumber \\ && + \Big({4u^3 \over x}+2x^2 -9-6u^2\Big) \ln[1+(x-u)^2] +4ux
(2-u^2) \nonumber \\ && +{1\over x}\bigg[ (7+15u^2-15x^2) [ \arctan(u+x)-
\arctan(x- u)] +{12 u^5 \over 5}-{21 u\over 2}   \nonumber \\ && -8u^3
\ln(1+4u^2) -\Big( 15u^2+{7 \over 4}\Big) \arctan 2u \bigg] + 3\theta(\sqrt{2}
u-x) \int_{y_{\rm min}}^{u/x} \!\!dy(x^2y^2-u^2) \nonumber \\ && \times {\cal 
A}_y \bigg[ {2s^2+s^4 \over 1+s^2} -2\ln(1+s^2) \bigg] + 
\int_{y_{\rm min}}^{1} \!\!dy \, {\cal A}_y \bigg[ {3 s^4 x^2(y^2-1) \over 
1+s^2}  -{9 s^4 \over 2}\nonumber \\ && +10 x y ( 3 \arctan s-3s +s^3 ) 
+(9+6u^2-6x^2y^2) [s^2-\ln(1+s^2)] \bigg] \Bigg\} \,, \end{eqnarray}  
with $x = p/m_\pi$ and the auxiliary functions $y_{\rm min}=\sqrt{1-u^2/x^2}$ 
and $s= xy+\sqrt{u^2-x^2+x^2y^2}$. In order to keep the notation compact we
have introduced the antisymmetrization prescription ${\cal A}_y[f(y)]=
f(y)-f(-y)$. Note that there is a term in eq.(15) which vanishes identically
above $p =\sqrt{2}k_f$. A geometrical explanation for this non-smooth behavior
is that an orthogonal pair of vectors connecting the origin with two points
inside a sphere ceases to exist if the center of the sphere is displaced too
far from the origin (namely by more than $\sqrt{2}$ times the sphere
radius). The orthogonality of the (momentum difference) vectors is imposed
here by the non-relativistic on-mass-shell condition for a nucleon. The
combined two-body, three-body and four-body contributions to $W(p,k_f)$ from
the iterated $1\pi$-exchange Fock diagram read on the other hand (for $p>
k_f$):   
\begin{eqnarray} W(p,k_f)^{(F)} & =& {\pi g_A^4M m_\pi^4 \over (4\pi f_\pi)^4}
\Bigg\{ u^3 x +{u^5 \over 5x} +{3 \over 2x} \int_{(x-u)/2}^{(x+u)/2}\!\! d\xi
\Big[(2\xi-x)^2-u^2\Big] {1+4\xi^2 \over 1+2\xi^2}\, \ln( 1+4\xi^2) \nonumber
\\ && +   {3 \over 4\pi}\theta(\sqrt{2}u-x) \int_{y_{\rm min}}^{1}\!\! dy
\int_{y_{\rm min}}^{1}\!\! dz \,{\theta(1-y^2-z^2) \over \sqrt{1-y^2-z^2}} \, 
{\cal A}_y \Big[ s^2-\ln(1+s^2) \Big] \nonumber \\ && \times {\cal A}_z \Big[
t^2-\ln(1+t^2) \Big] +{3\over x} \int_{-1}^1\!\! dy \int_0^u\!\! d\xi \, \xi^2 
\Big[\ln(1+\sigma^2)  -\sigma^2 \Big] \Big( 1- {1\over R} \Big) \Bigg\} \,, 
\end{eqnarray}  
with some new auxiliary functions $t= xz+\sqrt{u^2-x^2+x^2z^2}$ and $\sigma= 
\xi y+ \sqrt{u^2-\xi^2+\xi^2y^2}$ and $R=\sqrt{(1+x^2-\xi^2)^2+4\xi^2(1-y^2)}$.
It is also interesting to consider the imaginary single-particle potential
$W(p,k_f)$ in the chiral limit $m_\pi = 0$. One finds the following closed form
expressions:
\begin{equation} W(p,k_f)|_{m_\pi = 0} = {3\pi g_A^4 M \over (4\pi f_\pi)^4} 
\bigg\{ {7 k_f^5 \over 5p} -k_f^3 p -{2 \over 5p}(2k_f^2-p^2)^{5/2} \, 
\theta(\sqrt 2 k_f -p) \bigg\} \,, \qquad p> k_f \,,
\end{equation} 
\begin{equation} W(p,k_f)|_{m_\pi = 0} = {9\pi g_A^4 M \over 4(4\pi f_\pi)^4}
\,(k_f^2-p^2)^2 \,, \qquad p< k_f \,,\end{equation}
to which the iterated $1\pi$-exchange Hartree and Fock diagrams have
contributed in the ratio $4:-1$. The analytical results in eqs.(17,18) agree 
with Galitskii's calculation \cite{galitski} of a contact-interaction to 
second order. In the chiral limit $m_\pi =0$ the spin-averaged product of two 
$\pi NN$-interaction vertices gets canceled by the pion propagators and thus 
one is effectively dealing with a zero-range NN-contact interaction at second
order. The agreement with Galitskii's result \cite{galitski} serves as an 
important check on the technically involved calculation behind eqs.(15,16).  

\begin{figure}
\begin{center}
\includegraphics[scale=0.55,clip]{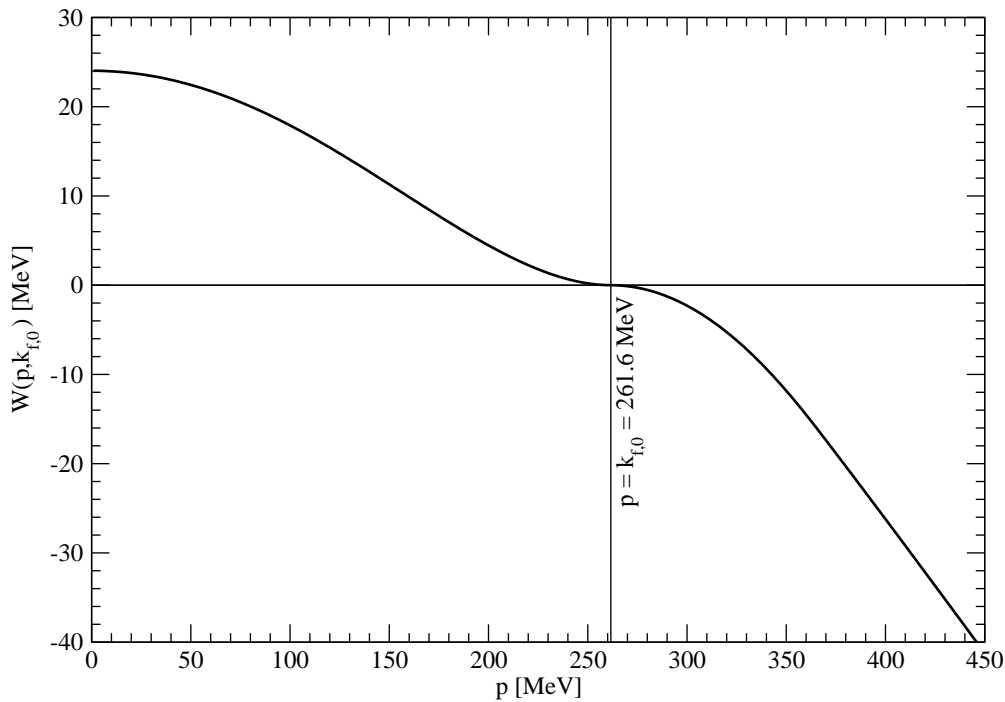}
\end{center}\vspace{-0.2cm}
\caption{The imaginary part of the single-particle potential $W(p,k_{f0})$ at 
saturation density $k_{f0}= 261.6\,$MeV as a function of the nucleon momentum 
$p$. The quadratic behavior around the Fermi surface $p = k_{f0}$ with a sign 
change of the curvature is required by Luttinger's theorem \cite{luttinger}.} 
\end{figure}

Fig.\,6 shows the imaginary part of the single-particle potential
$W(p,k_{f0})$ at saturation density $k_{f0}= 261.6\,$MeV as a function of the
nucleon momentum $p$. The quantity $\pm 2W(p,k_f)$ determines the width of a 
hole-state or a particle-state of momentum $p <k_f$ or $p >k_f$. The finite 
life time of these states originates from redistributing energy into 
additional particle-hole excitations. Our predicted value $W(0,k_{f0})=
24.0\,$MeV at $p=0$ lies in between the results $W(0,k_{f0})\simeq 20 \,$MeV of
ref.\cite{schuck}  employing the Gogny D1 effective interaction and
$W(0,k_{f0})\simeq 40 \,$MeV of ref.\cite{grange} using the Paris
NN-potential. As a consequence  of the decreasing phase-space for
redistributing the hole-state energy, the curve in Fig.\,6 drops with momentum
$p$ and $W(p,k_{f0})$ reaches zero at the Fermi surface $p= k_{f0}$. According
to Luttinger's theorem \cite{luttinger} this vanishing is of quadratic order 
$\sim (p-k_f)^2$, a feature which is clearly exhibited by the curve in
Fig.\,6. 

When crossing the Fermi surface the curvature of the imaginary
single-particle potential $W(p,k_f)$ flips the sign. From there on a rapid 
fall to negative values sets in. In fact the width $\Gamma_{sp} = -2 W(p,k_f)$
represents the spreading of a single-particle state above the Fermi surface
into two-particle-one-hole states with growing phase space as $p-k_f$ 
increases. The range of validity of the present chiral perturbation theory
calculation is again expected to be $p < 400\,$MeV. The rapid growth of
$\Gamma_{sp}$  beyond such limited momentum scales is not shared by the
results of refs.\cite{grange,schuck}.

\section{Nuclear energy density functional}
The energy density functional is a general starting point for
(non-relativistic) nuclear structure calculations within the framework of the
self-consistent mean-field approximation \cite{reinhard}. In this context
effective Skyrme forces \cite{sk3,sly,pearson}  have gained much popularity
because of their analytical simplicity and their ability to reproduce nuclear
properties over the whole periodic table. In a recent work \cite{efun} we have
calculated the nuclear energy density functional which emerges from (leading
and next-to-leading order) chiral pion-nucleon dynamics. The calculation in
ref.\cite{efun} included (only) the $1\pi$-exchange Fock-diagram and the
iterated $1\pi$-exchange Hartree and Fock diagrams. These few components alone
already lead to a good nuclear matter equation of state $\bar E(k_f)$. 
Therefore the interest here is on the additional effects from $2\pi$-exchange
with virtual $\Delta$-excitation contributing one order higher in the small
momentum expansion. Going up to quadratic order in spatial gradients
(i.e. deviations from homogeneity) the energy density functional relevant for
N=Z even-even nuclei reads \cite{efun}:      
\begin{equation} {\cal E}[\rho,\tau] = \rho\,\bar E(k_f)+\bigg[\tau-
{3\over 5} \rho k_f^2\bigg] \bigg[{1\over 2M}-{5k_f^2 \over 56 M^3}+F_\tau(k_f)
\bigg] + (\vec \nabla \rho)^2\, F_\nabla(k_f)\,,\end{equation}
with $\rho(\vec r \,) = 2k_f^3(\vec r\,)/3\pi^2$ the local nucleon density
(expressed in terms of a local Fermi momentum $k_f(\vec r\,)$) and $\tau
(\vec r\,)$ the local kinetic energy density. We have left out in eq.(19) the 
spin-orbit coupling term since the corresponding results (for $\Delta$-driven 
$2\pi$-exchange three-body spin-orbit forces\footnote{Interestingly, this 
three-body spin-orbit coupling is not a relativistic effect but independent of
the nucleon mass $M$.}) can be found in ref.\cite{3bodyso}. In 
phenomenological Skyrme parameterizations, the strength function $F_\tau(k_f)$ 
goes linearly with the density $\rho$, while $F_\nabla(k_f)$ is constant. The 
starting point for the construction of an explicit nuclear energy density 
functional ${\cal E}[\rho,\tau]$ is the bilocal density-matrix as given by a 
sum over the occupied energy eigenfunctions: $\sum_{\alpha\in \rm occ}
\Psi_\alpha( \vec r -\vec a/2)\Psi_\alpha^\dagger(\vec r +\vec a/2)$. 
According to Negele and Vautherin \cite{negele} it can be expanded in relative 
and center-of-mass coordinate, $\vec a$ and $\vec r$, with expansion 
coefficients determined by purely local quantities (nucleon density, kinetic 
energy density and spin-orbit density). As outlined in section 2 of 
ref.\cite{efun} the Fourier transform of the (so-expanded) density matrix 
defines in momentum space a ''medium insertion'' for the inhomogeneous  
many-nucleon system:  
\begin{equation} \Gamma(\vec p,\vec q\,) =\int d^3 r \, e^{-i \vec q \cdot
\vec r}\,\theta(k_f-|\vec p\,|) \bigg\{1 +{35 \pi^2 \over 8k_f^7}(5\vec
p\,^2 -3k_f^2) \bigg[ \tau - {3\over 5} \rho k_f^2 - {1\over 4} \vec \nabla^2 
\rho \bigg] \bigg\}\,. \end{equation}
The strength function $F_\tau(k_f)$ in eq.(19) emerges via a perturbation on 
top of the density of states $\theta(k_f-|\vec p\,|)$. As a consequence of 
that, $F_\tau(k_f)$ can be directly expressed in terms of the real 
single-particle potential $U(p,k_f)$ as:
\begin{equation} F_\tau(k_f) = {35 \over 4k_f^7} \int_0^{k_f} dp\,
p^2(5p^2-3k_f^2)\, U(p,k_f) \,. \end{equation}
Note that any $p$-independent contribution, in particular the $B_3$-term in 
eq.(9) and the $\zeta$-term in eq.(11), drops out. In the medium insertion 
eq.(20) $\tau-3\rho k_f^2/5$ is accompanied by $-\vec \nabla^2\rho/4$. After
performing a partial integration one is lead to the following decomposition:  
\begin{equation}F_\nabla(k_f) = {\pi^2 \over 8 k_f^2}\, {\partial F_\tau(k_f) 
\over  \partial k_f} +F_d(k_f) \,,\end{equation}
where $F_d(k_f)$ comprises all those contributions for which the $(\vec \nabla
\rho)^2$-factor originates directly from the momentum-dependence of the  
interactions. 

We enumerate now the contributions to the strength functions $F_{\tau,d}(k_f)$ 
generated by $2\pi$-exchange with virtual $\Delta$-excitation. We start with
(regularization dependent) contributions encoded in subtraction constants:
\begin{equation} F_\tau(k_f)^{(ct)}= B_5 {5 k_f^3 \over 3M^4} \,, \qquad
F_d(k_f)^{(ct)}=  {B_d \over M^4} \,, \end{equation}
where the new parameter $B_d =- M^4V''_C(0)/4$ stems from two-body Hartree 
diagrams and the momentum transfer dependence of the isoscalar central
NN-amplitude $V_C(q)$. Two-body Fock diagrams contribute only to $F_\tau(k_f)$ 
via a (subtracted) dispersion integral:    
\begin{eqnarray} F_\tau(k_f)^{(2F)} &=&{35 \over 24\pi^3 k_f^4} \int_{2m_\pi}^{
\infty} d\mu\,{\rm Im}(V_C+3W_C+2\mu^2V_T+6\mu^2W_T)\bigg\{ {8k_f^7 \over 35 
\mu^3} - {\mu k_f^3 \over 3}-6\mu^3k_f \nonumber \\ && + {\mu^5 \over 4k_f} +5
\mu^4 \arctan{2k_f\over \mu}+{\mu^3 \over 16k_f^3}(24k_f^4-18k_f^2\mu^2-\mu^4)
\ln\bigg( 1+{4k_f^2 \over \mu^2} \bigg) \bigg\} \,. \end{eqnarray}
The evaluation of the (left) three-body Hartree diagram in Fig.\,2 leads to
the results: 
\begin{equation}F_\tau(k_f)^{(3H)}={35 g_A^4 m_\pi^4 \over\Delta(2\pi f_\pi)^4}
\bigg\{ {13 \over 4}-{5 \over 24u^2}+{u^2\over 9} -{35 \over 12u} \arctan 2u 
+\bigg( {5\over 96 u^4}+{3\over 4u^2} -{3\over 4}\bigg)\ln(1+4u^2) \bigg\} \,, 
\end{equation}
\begin{equation} F_d(k_f)^{(3H)} = {g_A^4 m_\pi\over 128\pi^2 \Delta f_\pi^4}
\bigg\{ 23 \arctan 2u -{7\over u}\ln(1+4u^2)-16u - {2u(3+16u^2)\over
3(1+4u^2)^2}\bigg\} \,. \end{equation}
Somewhat more involved is the evaluation of the (right) Fock diagram in 
Fig.\,2 for which we find: 
\begin{equation} F_\tau(k_f)^{(3F)}={35g_A^4 m_\pi^4 \over \Delta(8\pi
f_\pi)^4 u^7}\int_0^u \!\!dx\Big[ 2G_S(x,u) \widetilde G_S(x,u)
+G_T(x,u)\widetilde  G_T(x,u) \Big] \,, \end{equation}
\begin{eqnarray} \widetilde G_S(x,u) &=& {4ux\over 3}(6u^4 -22u^2 -45+30x^2
-10u^2x^2) \nonumber \\ && +4x(10+9u^2-15x^2)\Big[\arctan(u+x)
+\arctan(u-x)\Big] \nonumber \\ && + (35x^2+14u^2x^2-10x^4-5-9u^2-4u^4) 
\ln{1+(u+x)^2 \over  1+(u-x)^2} \,,\end{eqnarray}
\begin{eqnarray} \widetilde G_T(x,u) &=& {ux\over 12}(69u^4 +70u^2 -15) -{u x^3
\over 12}(45+31u^2) -{15 u x^5\over 4} \nonumber \\ &&  -{u \over 4x}(1+u^2)^2
(5+3u^2) +{[1+(u+x)^2][1+(u-x)^2] \over 16 x^2}\nonumber \\ && \times  
(5+8u^2+3u^4 -18u^2x^2+ 15x^4)\ln{1+(u+x)^2 \over  1+(u-x)^2} \,,\end{eqnarray}
\begin{eqnarray} F_d(k_f)^{(3F)} &=& {g_A^4 m_\pi\over \pi^2 \Delta (8f_\pi)^4}
\bigg\{ -{3+12u^2+26u^4+40u^6 \over u^5 (1+4u^2)} \ln(1+4u^2)\nonumber \\ && +
{3\over 8 u^7}(1+2u^2+8u^4)\ln^2(1+4u^2) + {2(3+6u^2+16u^4)\over u^3(1+4u^2)} 
\bigg\} \,, \end{eqnarray}
with $G_{S,T}(x,u)$ defined in eqs.(7,8). The strength functions $F_{\tau,d}
(k_f)$ are completed by adding to the terms in eqs.(23-30) the 
contributions from $1\pi$-exchange and iterated  $1\pi$-exchange written down
in eqs.(9,11,12,14,15,18,19,22,24,27,28) of ref.\cite{efun}. In order to be
consistent with the calculation of the energy per particle $\bar E(k_f)$ and 
the single-particle potential $U(p,k_f)$ we complete the $1\pi$-exchange
contribution by its relativistic $1/M^2$-correction: 
\begin{eqnarray} F_\tau(k_f)^{(1\pi)} &=& {g_A^2m_\pi^3 u^{-5}\over(32\pi 
f_\pi M)^2} \bigg\{ {280 \over 3 } u^6  -{15 \over 2}+2u (525-700u^2-
96u^4) \arctan 2u \nonumber \\ && -64 u^8+744 u^4-1777 u^2 +\bigg( 
1050u^2-77 +{15 \over 8u^2} \bigg)\ln(1+4u^2) \bigg\} \,.   \end{eqnarray}
The expression in eq.(19) multiplying the kinetic energy density $\tau(\vec
r\,)$ has the interpretation of a reciprocal density dependent effective
nucleon mass:  
\begin{equation} \widetilde M^*(\rho) = M \bigg[1-{5k_f^2 \over 28M^2}+ 2M\, 
F_\tau(k_f)\bigg]^{-1} \,. \end{equation}
We note that this effective nucleon mass $\widetilde M^*(\rho)$ (entering the
nuclear energy density functional) is conceptually different from the so-called
''Landau''-mass $M^*(k_f)$  defined in eq.(13). Only if the real 
single-particle potential has a simple quadratic dependence on the nucleon
momentum, $U(p,k_f)=U_0(k_f)+p^2U_1(k_f)$, do these two variants of effective 
nucleon mass agree with each other (modulo very small differences related to
the relativistic $(k_f/2M)^2$-correction). 
\begin{figure}
\begin{center}
\includegraphics[scale=0.55,clip]{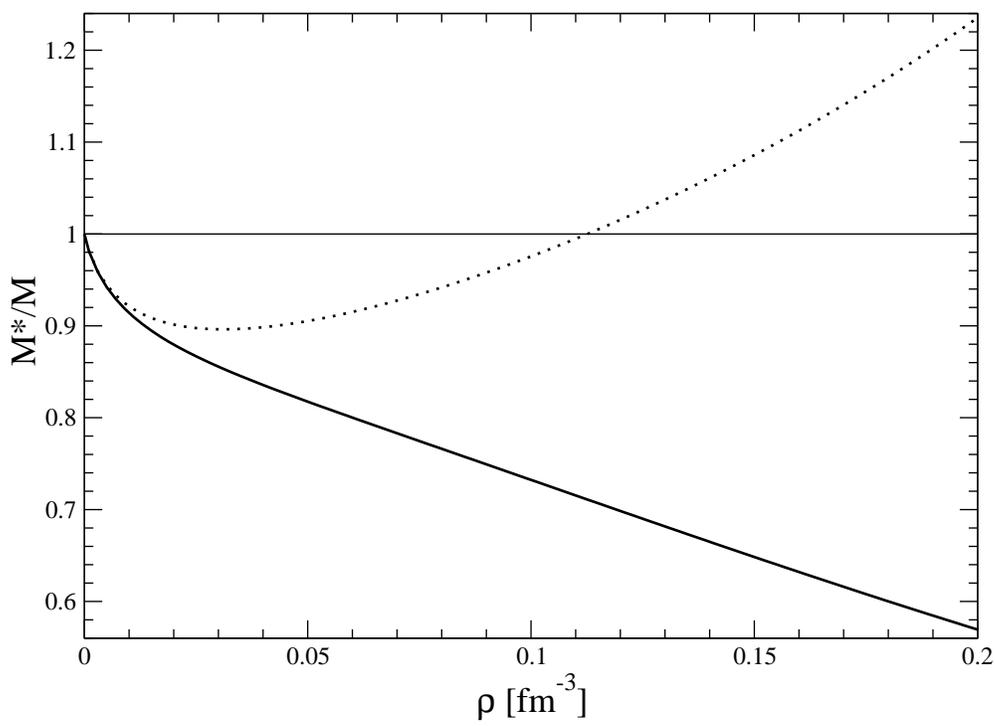}
\end{center}\vspace{-0.2cm}
\caption{The effective nucleon mass $\widetilde M^*(\rho)$ divided by the
free nucleon mass $M$ as a function of the nucleon density $\rho$. The dotted 
line shows the result of ref.\cite{efun} based on single and iterated 
pion-exchange only. The full line includes in addition the effects from
$2\pi$-exchange with virtual $\Delta$-excitation.}   
\end{figure}

In Fig.\,7 we show the ratio effective-over-free nucleon mass $\widetilde
M^*(\rho)/M$ as a function of the nucleon density $\rho= 2k_f^3/3\pi^2$. The 
dotted line corresponds to the result of ref.\cite{efun} based on $1\pi$- and 
iterated $1\pi$-exchange only. The full line includes in addition the effects 
from $2\pi$-exchange with virtual $\Delta$-excitation. It is clearly visible
that the inclusion of the $\pi N \Delta$-dynamics leads to substantial
improvement of the effective nucleon mass $\widetilde M^*(\rho)$ since now it 
decreases monotonically with the density. This behavior is of course a 
direct reflection of the improved momentum dependence of the real 
single-particle potential $U(p,k_f)$ (see Fig.\,4). Our prediction for the 
effective nucleon mass at saturation density, $\widetilde M^*(\rho_0) = 0.64M$,
is comparable to the typical value $\widetilde M^*(\rho_0) \simeq 0.7M$ of
phenomenological Skyrme forces \cite{sk3,sly}. The full curve in Fig.\,7 
displays another interesting and important feature, namely strong curvature 
effects at low densities $\rho < 0.05\,$fm$^{-3}$. They originate from the
explicit presence of the small mass scale $m_\pi = 135\,$MeV in our
calculation.    

\begin{figure}
\begin{center}
\includegraphics[scale=0.55,clip]{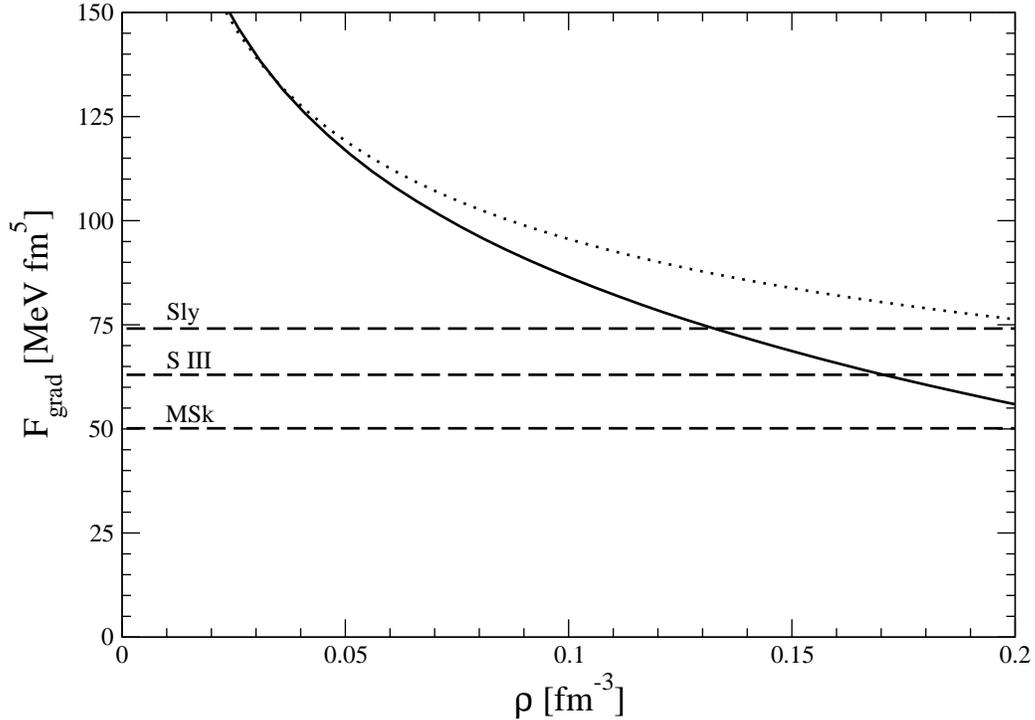}
\end{center}\vspace{-0.2cm}
\caption{The strength function $F_\nabla(k_f)$ multiplying the $(\vec \nabla
\rho)^2$-term in the nuclear energy density functional versus the nucleon 
density $\rho= 2k_f^3/3\pi^2$. The dotted line shows the result of  
ref.\cite{efun} based on single and iterated pion-exchange only. The full 
line includes in addition the effects from $2\pi$-exchange with virtual 
$\Delta$-excitation.}   
\end{figure}

Fig.\,8 shows the strength function $F_\nabla(k_f)$ belonging to the $(\vec 
\nabla \rho)^2$-term in the nuclear energy density functional versus the
nucleon density $\rho= 2k_f^3/3\pi^2$. The dotted line gives the result of  
ref.\cite{efun} based on $1\pi$- and iterated $1\pi$-exchange only and the 
full line includes in addition the effects from $2\pi$-exchange with virtual 
$\Delta$-excitation. The subtraction constant $B_d$ (representing density 
independent short-range contributions) has been set to zero. In the region
around saturation density $\rho_0 \simeq 0.16\,$fm$^{-3}$ one observes a clear 
improvement since there the full line meets the band spanned by the three 
phenomenological Skyrme forces SIII \cite{sk3}, Sly \cite{sly} and MSk
\cite{pearson}. The strong rise of $F_\nabla(k_f)$ towards low densities
remains however. As explained in ref.\cite{efun} this has to do with chiral
singularities (of the form $m^{-2}_\pi$ and $m^{-1}_\pi$) in the contributions
from $1\pi$-exchange and iterated $1\pi$-exchange.     

The knowledge of the strength function $F_\nabla(k_f)$ and the equation of 
state $\bar E(k_f)$ allows one to calculate the surface energy of 
semi-infinite nuclear matter \cite{brack} via: 
\begin{equation} a_s = 2(36\pi \rho_0^{-2})^{1/3} \int_0^{\rho_0} 
\!\!d\rho \sqrt{ \rho \,F_\nabla(k_f)[\bar E(k_f)-\bar E_0]}\,. \end{equation} 
Here, we have inserted into the formula for the surface energy $a_s$ (see 
eq.(5.32) in ref.\cite{brack}) that density profile $\rho(z)$ which minimizes 
it. Numerical evaluation of eq.(33) gives $a_s= 24.2\,$MeV. This number
overestimates semi-empirical determinations of the surface energy, such as 
 $a_s = 20.7\,$MeV of ref.\cite{blaizot} or $a_s =18.2\,$MeV of 
ref.\cite{brack}, by $17\%$ or more.\footnote{One could reproduce the surface
energy $a_s = 20.7\,$MeV of ref.\cite{blaizot} by adjusting the short-range
parameter $B_d$ in eq.(32) to the value $B_d = -75$. The full curve for
$F_\nabla(k_f)$ in Fig.\,8 would then be shifted downward by 29\,MeVfm$^5$. 
Compared to $B_3 = -8$ the fitted number $B_d = -75$ seems to be rather 
large.} The reason for our high value  $a_s=24.2\,$MeV is of course the strong 
rise of the strength function $F_\nabla(k_f)$ at low densities. Its
derivation is based on the density-matrix expansion of Negele and Vautherin
\cite{negele} which has been found to become inaccurate at low and non-uniform
densities \cite{dobacz}. Therefore one should not trust the curves in Fig.\,8
below $\rho = 0.05\,$fm$^{-3}$. Getting the right order of magnitude for
$F_\nabla(k_f)$ in the density region $0.1\,$fm$^{-3}<\rho <0.2\,$fm$^{-3}$ is
already a satisfactory result. 

\section{Nuclear matter at finite temperatures}
In this section, we discuss nuclear matter at finite temperatures $T\leq 
30\,$MeV. We are particularly interested in the first-order liquid-gas phase 
transition of isospin-symmetric nuclear matter and its associated critical 
point $(\rho_c,T_c)$. As outlined in ref.\cite{liquidgas} a thermodynamically 
consistent extension of the present (perturbative) calculational scheme to 
finite temperatures is to relate it directly to the free energy per particle 
$\bar F(\rho,T)$, whose natural thermodynamical variables are the nucleon 
density $\rho$ and the temperature $T$. In that case the free energy density 
$\rho\bar F(\rho,T)$ of isospin-symmetric nuclear matter consists of a sum of
convolution integrals over interaction kernels ${\cal K}_j$ multiplied by
powers of the density of nucleon states in momentum space:\footnote{Since the
temperature $T$ is comparable to an average kinetic energy we count $T$ of
quadratic order in small momenta.}
\begin{equation} 
d(p_j) = {p_j\over 2\pi^2} \bigg[ 1+\exp{p_j^2 -2M \tilde \mu
\over 2M T} \bigg]^{-1} \,. \end{equation}
The effective one-body ''chemical potential" $\tilde \mu(\rho,T)$ entering the
Fermi-Dirac distribution in eq.(34) is determined by the relation to the 
particle density $\rho = 4\int_0^\infty d p_1\,p_1 d(p_1)$. We summarize now
the additional interaction kernels arising from $2\pi$-exchange with virtual 
$\Delta$-excitation. The two-body kernels read:
\begin{equation} {\cal K}_2^{(ct)}= 24\pi^2 B_3 {p_1p_2 \over M^2} +20\pi^2 B_5
{p_1p_2 \over M^4}(p_1^2+p_2^2)\,, \end{equation}
\begin{eqnarray} {\cal K}_2^{(F)} &=& {1 \over \pi} \int_{2m_\pi}^{
\infty}\!\! d\mu\,{\rm Im}(V_C+3W_C+2\mu^2V_T+6\mu^2W_T)\nonumber \\ && \times
\bigg\{\mu  \ln{\mu^2+(p_1+p_2)^2 \over \mu^2+(p_1-p_2)^2} -{4p_1p_2\over \mu}
+{4p_1p_2\over \mu^3}(p_1^2+ p_2^2) \bigg\} \,.  \end{eqnarray}
Note that the $B_3$-term in eq.(35) generates a temperature independent 
contribution to the free energy per particle, $\bar F(\rho,T)^{(B_3)}= 3\pi^2 
B_3\rho/2M^2$, which drops linearly with density. Temperature and density
dependent Pauli blocking effects are incorporated in the three-body kernel
${\cal K}_3$.  The contributions of the Hartree and Fock diagrams in Fig.\,2
to the  three-body kernel read: 
\begin{equation} {\cal K}^{(H)}_3 = {3g_A^4 p_3 \over \Delta f_\pi^4} \bigg\{
2p_1p_2 (1+\zeta) +{2m_\pi^4p_1p_2\over [m_\pi^2+(p_1+p_2)^2][m_\pi^2+(p_1-p_2
)^2]} - m_\pi^2 \ln{m_\pi^2+(p_1+p_2)^2 \over m_\pi^2+(p_1-p_2)^2} \bigg\}
\,, \end{equation}
\begin{equation} {\cal K}^{(F)}_3 = -{g_A^4 \over 4\Delta f_\pi^4p_3} \Big[
2X(p_1)X(p_2)+ Y(p_1)Y(p_2)\Big] \,, \end{equation}
\begin{equation} X(p_1) = 2p_1p_3 -{m_\pi^2\over 2} \ln{m_\pi^2+(p_1+p_3)^2 
\over m_\pi^2+(p_1-p_3)^2} \,, \end{equation}
\begin{eqnarray}Y(p_1) &=& {p_1\over 4p_3}(5p_3^2-3m_\pi^2-3p_1^2) \nonumber \\
&& +{3(p_1^2-p_3^2+m_\pi^2)^2+4m_\pi^2p_3^2 \over 16 p_3^2} 
\ln{m_\pi^2+(p_1+p_3)^2 \over m_\pi^2+(p_1-p_3)^2} \,.   \end{eqnarray}
The remaining kernels building up the free nucleon gas part and the interaction
contributions from $1\pi$-exchange and iterated $1\pi$-exchange have been 
written down in eqs.(4,5,6,7,11,12) of ref.\cite{liquidgas}. It is needless to
say that the extension of our nuclear matter calculation to finite
temperatures $T$ does not introduce any new adjustable parameter. 

\begin{figure}
\begin{center}
\includegraphics[scale=0.55,clip]{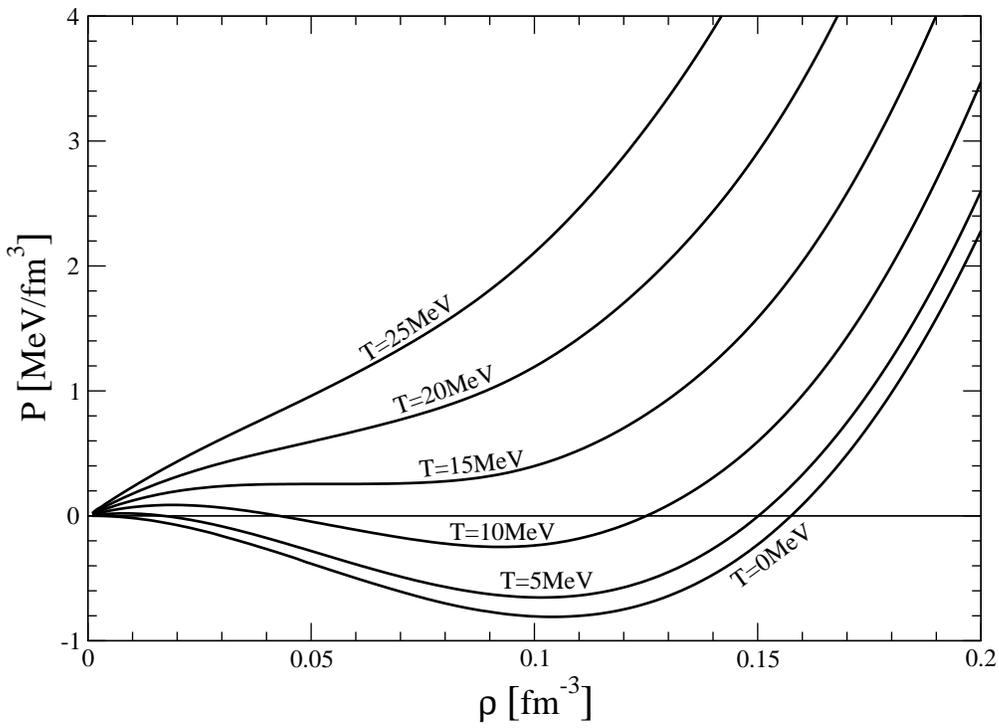}
\end{center}\vspace{-0.2cm}
\caption{Pressure isotherms $P(\rho,T)= \rho^2 \partial \bar F(\rho,T)/\partial
\rho$ of isospin-symmetric nuclear matter at finite temperature $T$. The 
coexistence region of the liquid and the gas phase ends at the critical point:
$\rho_c \simeq 0.053\,$fm$^{-3}$, $T_c \simeq 15\,$MeV.}
\end{figure}

Fig.\,9 shows the calculated pressure isotherms $P(\rho,T) = \rho^2 \partial
\bar F(\rho,T)/\partial \rho$ of isospin-symmetric nuclear matter at six 
selected temperatures $T=0,\,5,\,10,\,15,\,20,\,25\,$MeV. As it should be these
curves display a first-order liquid-gas phase transition similar to that of the
van-der-Waals gas. The coexistence region between the liquid and the gas phase
(which has to be determined by the Maxwell construction) terminates at the
critical temperature $T_c$. From there on the pressure isotherms $P(\rho,T)$
grow monotonically with the nucleon density $\rho$. We find here a critical
temperature of $T_c \simeq 15\,$MeV and a critical density of $\rho_c \simeq 
0.053\,$fm$^{-3}\simeq \rho_0/3$. This critical temperature is close to the  
value $T_c = (16.6\pm 0.9)\,$MeV extracted in ref.\cite{natowitz} from 
experimentally observed limiting temperatures in heavy ion collisions. In
comparison, a critical temperature of $T_c = (20\pm 3)\,$MeV has been 
extracted in ref.\cite{multifrag} from multi-fragmentation data in 
proton-on-gold collisions. Most other nuclear matter calculations find a 
critical temperature somewhat higher than our value, typically $T_c \simeq 18\,
$MeV \cite{urbana,sauer,kapusta}. The reduction of $T_c$ in comparison to 
$T_c \simeq 25.5\,$MeV obtained previously in ref.\cite{liquidgas} results
from the much improved momentum dependence of the real single-particle
potential $U(p,k_{f0})$ near the Fermi surface $p=k_{f0}$ (see Fig.\,4). As a
general rule the critical temperature $T_c$ grows with the effective nucleon
mass $M^*(k_{f0})$ at the Fermi surface. 

The single-particle properties around the Fermi surface are decisive for the 
thermal excitations and therefore they crucially influence the low temperature 
behavior of nuclear matter. The inclusion of the chiral $\pi N\Delta$-dynamics 
leads to a realistic value of the density of (thermally excitable) nucleon 
states at the Fermi surface. This is an important observation.    

\section{Equation of state of pure neutron matter}
This section is devoted to the equation of state of pure neutron matter. In
comparison to the calculation of isospin-symmetric nuclear matter in section 2
only the isospin factors of the $2\pi$-exchange diagrams with virtual
$\Delta$-excitation change. We summarize now the contributions to the energy 
per particle $\bar E_n(k_n)$ of pure neutron matter. The two-body terms read: 
\begin{equation} \bar E_n(k_n)^{(ct)}= B_{n,3} {k_n^3 \over M^2} + B_{n,5}
{k_n^5  \over  M^4}\,,  \end{equation}
\begin{eqnarray} \bar E_n(k_n)^{(2F)}&=& {1 \over 8\pi^3} 
\int_{2m_\pi}^{\infty}\!\! d\mu \,{\rm Im}(V_C+W_C+2\mu^2V_T+2\mu^2W_T)\bigg\{
3\mu k_n -{4k_n^3 \over 3\mu }\nonumber \\ &&+{8k_n^5 \over 5\mu^3 }
-{\mu^3\over 2k_n}-4\mu^2 \arctan{2k_n\over\mu}  +{\mu^3 \over 8k_n^3}(12k_n^2
+\mu^2) \ln\bigg( 1+{4k_n^2 \over \mu^2} \bigg) \bigg\} \,, \end{eqnarray}
with $B_{n,3}$ and $B_{n,5}$ two new subtraction constants. The relative 
weights of isoscalar ($V_{C,T}$) and isovector ($W_{C,T}$) NN-amplitudes have 
changed by a factor 3 in comparison to eq.(2). The three-body terms generated
by the Hartree and Fock diagrams in Fig.\,2 read:  
\begin{equation} \bar E_n(k_n)^{(3H)}={g_A^4 m_\pi^6 \over 6\Delta(2\pi f_\pi
)^4}  \bigg[ {2\over3}u^6 + u^2-3u^4+5u^3 \arctan2u-{1\over 4}(1+9u^2)
\ln(1+4u^2) \bigg] \,, \end{equation}  
\begin{equation} \bar E_n(k_n)^{(3F)}=-{g_A^4 m_\pi^6  u^{-3}\over 4\Delta(4\pi
f_\pi)^4 }\int_0^u\!\! dx\Big[ G^2_S(x,u)+2G^2_T(x,u)\Big] \,, \end{equation}
with $G_{S,T}(x,u)$ defined in eqs.(7,8). We emphasize that in this section the
meaning of $u$ changes to $u= k_n/m_\pi$, where $k_n$ denotes the neutron 
Fermi momentum. All three three-body diagrams in Fig.\,2 have now the same 
isospin factor $2/3$ since between neutrons only the $2\pi^0$-exchange is
possible. Note that the Pauli-exclusion principle forbids a three-neutron
contact-interaction and therefore eq.(43) is free of any $\zeta$-term.
The additional contributions to $\bar E_n(k_n)$ from the (relativistically
improved) kinetic energy, from $1\pi$-exchange and from iterated
$1\pi$-exchange have been written down in eqs.(32-37) of ref.\cite{nucmat}.   

\begin{figure}
\begin{center}
\includegraphics[scale=0.54,clip]{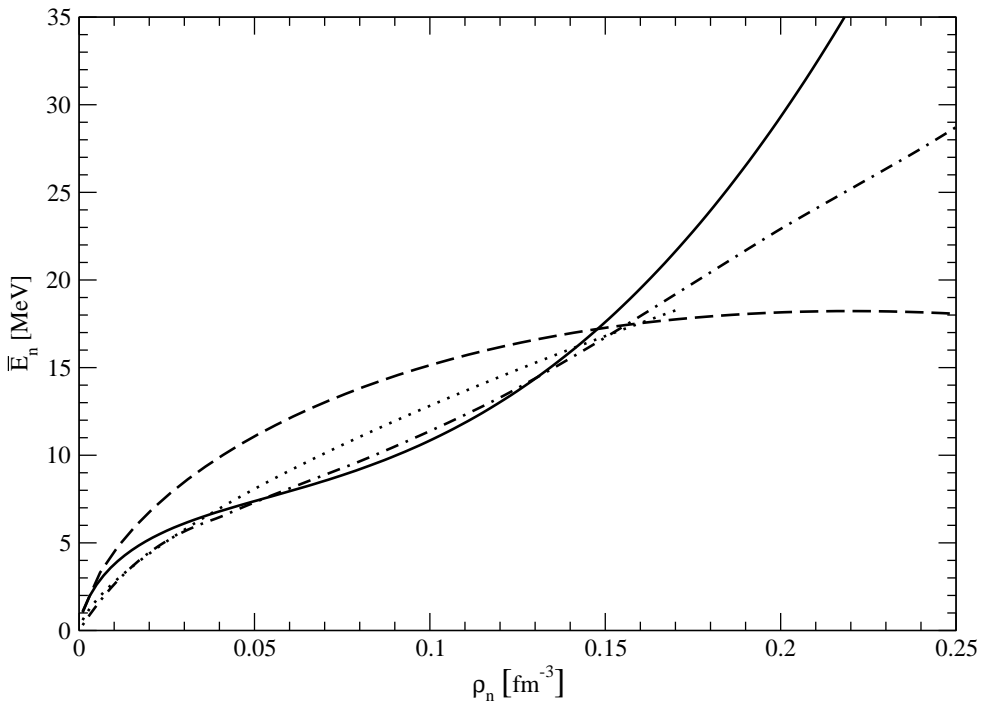}
\end{center}\vspace{-0.4cm}
\caption{The energy per particle $\bar E_n(k_n)$ of pure neutron matter as a
function of the neutron density $\rho_n = k_n^3/3\pi^2$. The dashed curve gives
the result of ref.\cite{nucmat}. The full curve includes the $\pi N\Delta
$-dynamics and two adjusted short-range parameters $B_{n,3}=-0.94$ and 
$B_{n,5}=-3.58$. The dashed-dotted curve stems from the sophisticated 
many-body calculation of the Urbana group \cite{akmal}. The dotted curve gives
one half of the kinetic energy $\bar E_{\rm kin}(k_n)/2 = 3k_n^2/20M$.}   
\end{figure}

Fig.\,10 shows the energy per particle $\bar E_n(k_n)$ of pure neutron matter 
as a function of the neutron density $\rho_n = k_n^3/3\pi^2$. The dashed
(concave) curve gives the result of our previous calculation in 
ref.\cite{nucmat}. The full curve includes the chiral $\pi N\Delta$-dynamics. 
The short-range parameters $B_{n,3}$ and $B_{n,5}$ (controlling the 
contribution of a nn-contact interaction to $\bar E_n(k_n)$) have been 
adjusted to the values $B_{n,3} = -0.95$ and $B_{n,5}=-3.58$.\footnote{The 
short-range parameters $B_{n,3}$ and $B_{n,5}$ have been adjusted such that 
the asymmetry energy at saturation density takes on the value $A(k_{f0})=
34\,$MeV.} The dashed-dotted curve in Fig.\,10 stems from
the sophisticated many-body calculation of the Urbana group \cite{akmal}, to
be considered as representative of realistic neutron matter calculations. 
Moreover, the dotted curve gives one half of the kinetic energy of a free 
neutron gas, $\bar E_{\rm kin}(k_n)/2 =3k_n^2/20M$. Results of recent quantum 
Monte-Carlo calculation in ref.\cite{montecarlo} have demonstrated that the 
neutron matter equation of state at low neutron densities $\rho_n
<0.05\,$fm$^{-3}$ is well approximated by this simple form. One observes that 
up to $\rho_n = 0.16\,$fm$^{-3}$ our result for $\bar E_n(k_n)$ is very close 
to that of the sophisticated many-body calculation \cite{akmal,montecarlo}. At
higher densities we find a stiffer neutron matter equation of state. Again, 
one should not expect that our approach works at Fermi momenta larger than 
$k_n =350\,$MeV (corresponding to $\rho_n = 0.19\,$fm$^{-3}$). One of the most
important results of the present calculation is that the unrealistic downward
bending of $\bar E_n(k_n)$ (as shown by the dashed curve in Fig.\,10)
disappears after the inclusion of the chiral $\pi N\Delta$-dynamics. This is a
manifestation of improved isospin properties.     
        
\section{Asymmetry energy}
As a further test of isospin properties we consider in this section the 
density dependent asymmetry energy $A(k_f)$. The asymmetry energy is generally 
defined by the expansion of the energy per particle of isospin-asymmetric 
nuclear matter (described by different proton and neutron Fermi momenta 
$k_{p,n} = k_f(1\mp \delta)^{1/3}$) around the symmetry line: 
\begin{equation} \bar E_{\rm as}(k_p,k_n) = \bar E(k_f) + \delta^2\, A(k_f) 
+ {\cal O}(\delta^4)\,. \end{equation} 
Following the scheme in the previous sections we summarize the contributions to
the asymmetry energy $A(k_f)$. The two-body terms read:   
\begin{equation} A(k_f)^{(ct)}= (2B_{n,3}-B_3){k_f^3 \over M^2} +  (3B_{n,5}
-B_5)  {10k_f^5 \over  9M^4}\,,  \end{equation}
\begin{eqnarray} A(k_f)^{(2F)}&=& {1 \over 12\pi^3} \int_{2m_\pi}^{\infty} \!\!
d\mu  \bigg\{{\rm Im}(V_C+2\mu^2V_T)\bigg[\mu k_f-{2k_f^3 \over \mu} +{16k_f^5
\over 3\mu^3} - {\mu^3 \over 4k_f}\ln\bigg( 1+ {4k_f^2 \over \mu^2} \bigg)
\bigg] \nonumber \\ && + {\rm Im}(W_C+2\mu^2W_T) \bigg[ 3\mu k_f+{2k_f^3 \over 
\mu}-{\mu \over 4k_f} ( 8k_f^2+3\mu^2) \ln\bigg( 1+{4k_f^2 \over \mu^2}\bigg)
\bigg]\bigg\}\,. \end{eqnarray}
In eq.(46) we have taken care of the fact that there are only two independent
(S-wave) NN-contact couplings which can produce terms linear in density. It is
surprising that also the other coefficient $10(3B_{n,5}-B_5)/9$ in front of
the $k_f^5/M^4$-term is completely fixed. This fact can be shown on the basis
of the most general order-$p^2$ NN-contact interaction written down in
eq.(6) of ref.\cite{ordonez}. Out of the seven low-energy constants
$C_1,\dots,C_7$ only two independent linear combinations, $C_2$ and
$C_1+3C_3+C_6$, come into play for homogeneous and spin-saturated nuclear
matter. The contribution of the three-body Hartree diagram in Fig.\,2  to the
asymmetry energy $A(k_f)$ has the following analytical form:  
\begin{equation} A(k_f)^{(3H)}={g_A^4 m_\pi^6 u^2\over 9\Delta(2\pi f_\pi)^4} 
\bigg[ \bigg({9\over 4}+4u^2\bigg)\ln(1+4u^2)-2u^4(1+3\zeta) -8u^2-{u^2\over 
1+4u^2} \bigg] \,, \end{equation}
with the abbreviation $u =k_f/m_\pi$. The parameter $\zeta = -3/4$ is again
related to the three-nucleon contact interaction $\sim (\zeta g_A^4/\Delta 
f_\pi^4)\, (\bar N N)^3$ which has the interesting property that it 
contributes equally but with opposite sign to the energy per particle $\bar
E(k_f)$ and the asymmetry energy $A(k_f)$. Furthermore, both three-body Fock
diagrams in Fig.\,2 add up to give rise a contribution to the asymmetry
$A(k_f)$ which can be represented as:    
\begin{eqnarray} A(k_f)^{(3F)}&=&{g_A^4 m_\pi^6 u^{-3}\over 36\Delta(4\pi 
f_\pi)^4} \int_0^u\!\! dx\Big\{ 4G_{S01}G_{S10}-2G_{S01}^2-6G_{S10}^2 
\nonumber \\ && +2 G_S (3G_S+8 G_{S01}-3 G_{S02}+2 G_{S11}-3 G_{S20})+2G_{T01}
G_{T10} \nonumber \\ && -7G_{T01}^2-3G_{T10}^2  + G_T(3G_T+8 G_{T01}-3G_{T02}
+2G_{T11} -3 G_{T20}) \Big\} \,.\end{eqnarray}
The auxiliary functions $G_{S,T}(x,u)$ have been defined in eqs.(7,8) and we
have introduced a double-index notation for their partial derivatives:  
\begin{equation} G_{Ijk}(x,u) = x^j u^k {\partial^{j+k} G_I(x,u) \over 
\partial x^j \partial u^k} \,, \qquad I=S,T \,, \quad 1\leq j+k \leq
2\,. \end{equation}
For notational simplicity we have omitting the arguments $x$ and $u$ in the
integrand of eq.(49). The asymmetry energy $A(k_f)$ is completed by adding to
the terms in eqs.(46-49) the contributions from the (relativistically
improved) kinetic energy, $1\pi$-exchange and iterated $1\pi$-exchange written
down in eqs.(20-26) of ref.\cite{nucmat}. 

\begin{figure}
\begin{center}
\includegraphics[scale=0.55,clip]{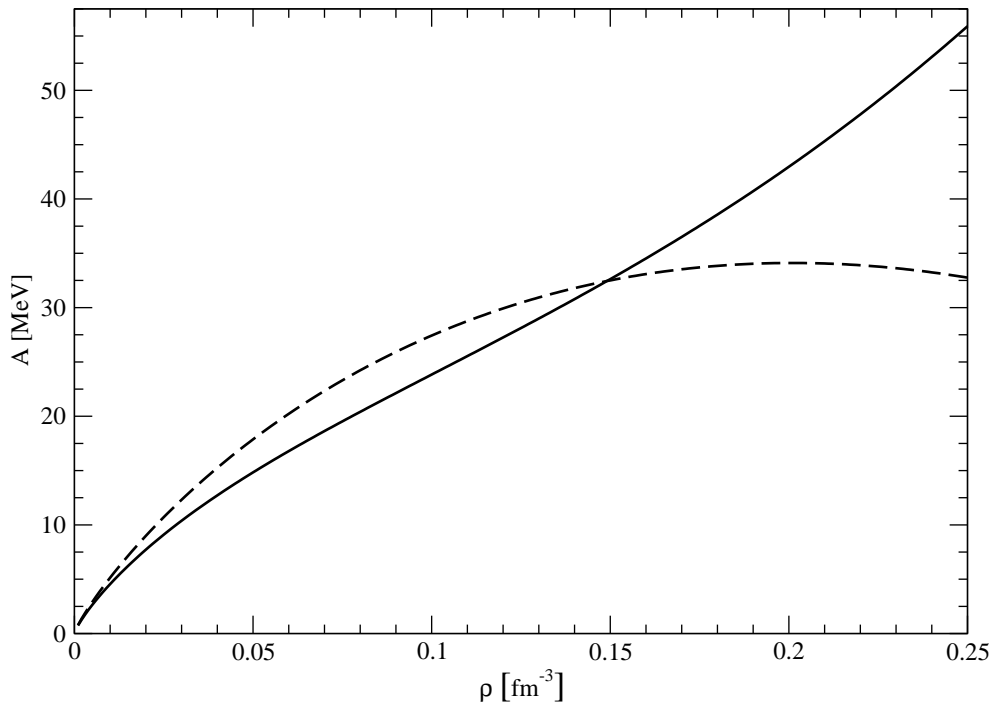}
\end{center}\vspace{-0.2cm}
\caption{The asymmetry energy $A(k_f)$ as a function of the nucleon density 
$\rho= 2k_f^3/3\pi^2$. The dashed curve shows the result of ref.\cite{nucmat}.
The full curve includes the chiral $\pi N \Delta$-dynamics.}
\end{figure}

In the calculation of the asymmetry energy we use consistently the previously 
fixed short-distance parameters $B_3 = -7.99$ and $B_{n,3}=-0.95$, as well as 
$B_5 = 0$ and $B_{n,5}=-3.58$. Fig.\,11 shows the asymmetry energy $A(k_f)$ as
a function of the nucleon density $\rho= 2k_f^3/3\pi^2$. The dashed (concave)
curve  corresponds to the result of ref.\cite{nucmat}. The full curve includes
the chiral $\pi N\Delta$-dynamics. The corresponding value of the asymmetry 
energy at saturation density $\rho_0=0.157\,$fm$^{-3}$ is $A(k_{f0})= 
34.0\,$MeV. It decomposes as $A(k_{f0})= (12.1+119.3-109.9 +12.5)\,$MeV into
contributions of second, third, fourth and fifth power of small momenta, again
with a balance between large third and fourth order terms \cite{nucmat}. 
The value $A(k_{f0})= 34.0\,$MeV is consistent with most of the existing 
empirical determinations of the asymmetry energy. For example, a recent
microscopic  estimate in a relativistic mean-field model (constrained by some
specific properties of certain nuclei) gave the value $A(k_{f0})= (34\pm2)
\,$MeV \cite{dario}. For comparison, other empirical values obtained from
extensive fits of nuclide masses are $A(k_{f0}) = 36.8\,$MeV \cite{blaizot} or 
$A(k_{f0})= 33.2\,$MeV \cite{seeger}. The slope of the asymmetry energy at 
saturation density, $L=k_{f0}A'(k_{f0})$, is likewise an interesting quantity.
As demonstrated in Fig.\,11 of ref.\cite{furn} the neutron skin thickness of 
$^{208}$Pb is linearly correlated with the slope parameter $L$. We extract 
from the full curve in Fig.\,11 the value $L=90.8\,$MeV. This prediction is 
not far from $L\simeq 100\,$MeV quoted in ref.\cite{blaizot} and $L=119.2\, 
$MeV obtained from the ''standard'' relativistic force NL3 \cite{nl3}. 
Furthermore, we extract from the curvature of our asymmetry energy $A(k_f)$ at
saturation density $\rho_0$ the positive asymmetry compressibility $K_{\rm
as}= k_{f0}^2 A''(k_{f0})-2L = 160.5\,$MeV.  

Again, the most important feature visible in Fig.\,11 is that the inclusion 
of the chiral $\pi N\Delta$-dynamics eliminates the (unrealistic) downward
bending of the asymmetry $A(k_f)$ at higher densities $\rho>0.2\,$fm$^{-3}$
(as displayed by the dashed curve in Fig.\,11). This is once more a 
manifestation of improved isospin properties.

\section{Isovector single-particle potential}
In this section we generalize the calculation of the single-particle potential
to isospin-asymmetric (homogeneous) nuclear matter. Any relative excess of
neutrons over protons in the nuclear medium leads to a different
''mean-field'' for a proton and a neutron. This fact is expressed by the
following decomposition of the (real) single-particle potential in
isospin-asymmetric nuclear matter:
\begin{equation} U(p,k_f) - U_I(p,k_f) \, \tau_3\,\delta +{\cal O}(\delta^2)\,,
\qquad \delta = {\rho_n - \rho_p \over \rho_n + \rho_p}\,, \end{equation}
with $U(p,k_f)$ the isoscalar (real) single-particle potential discussed in
section 3. The term linear in the isospin-asymmetry parameter $\delta=(\rho_n 
- \rho_p)/(\rho_n + \rho_p)$ defines the (real) isovector single-particle 
potential $U_I(p,k_f)$, and $\tau_3 \to \pm 1$ for a proton or a neutron. 
Without going into further technical details we summarize now the 
contributions to $U_I(p,k_f)$. The two-body terms read:
\begin{equation} U_I(p,k_f)^{(ct)}= (2B_{n,3}-B_3){2k_f^3 \over M^2}+(2B_{n,5}
-B_5) {5k_f^3 \over  3M^4}(k_f^2+p^2)\,,  \end{equation}
\begin{eqnarray}U_I(p,k_f)^{(2F)}&=&{k_f^2\over 12\pi^3}\int_{2m_\pi}^{\infty} 
\!\! d\mu \,{\rm Im}(V_C-W_C+2\mu^2V_T-2\mu^2W_T) \nonumber \\ && \times  
\bigg\{ {4k_f \over \mu^3} (k_f^2+p^2-\mu^2) +{\mu \over p}\ln{\mu^2+(k_f+p)^2 
\over \mu^2+(k_f-p)^2}\bigg\}\,. \end{eqnarray}
The contribution of the three-body Hartree diagram in Fig.\,2 and the
three-body contact term can be written in analytical form: 
\begin{equation} U_I(p,k_f)^{(3H)}= {2g_A^4 m_\pi^6 u^5 \over 9 \Delta(2\pi 
f_\pi)^4} \bigg\{ {1\over x} \ln{1+(u+x)^2 \over 1+(u-x)^2}- {2u \over 
[1+(u+x)^2] [1+(u-x)^2]} -2u(1+3 \zeta) \bigg\}\,, \end{equation} 
and the total contribution of both three-body Fock diagrams in Fig.\,2 can be
represented as: 
\begin{eqnarray} U_I(p,k_f)^{(3F)}&=&{g_A^4m_\pi^6u x^{-2}\over 18 \Delta 
(4\pi f_\pi)^4}\bigg\{ 2 \,G_S(x,u)\, {\partial G_S(x,u) \over \partial u}
+G_T(x,u) \,{\partial G_T(x,u) \over \partial u} \nonumber \\ && +2\, G_S(u,u) 
\, {\partial G_S(u,x) \over \partial x} +G_T(u,u)\, {\partial G_T(u,x) \over 
\partial x} \nonumber \\ && - \int_0^u \!\!d\xi \bigg[ 2\,{\partial G_S(\xi,u) 
\over \partial u}\, {\partial G_S(\xi,x)  \over \partial x}+ 7 \,{\partial 
G_T(\xi,u) \over \partial u}\, {\partial G_T(\xi,x)  \over \partial x} \bigg]
\bigg\} \,, \end{eqnarray}
where $x= p/m_\pi$ and $u = k_f/m_\pi$. The auxiliary functions $G_{S,T}(x,u)$
have been defined in eqs.(7,8). The (real) isovector single-particle
potential $U_I(p,k_f)$ (restricted to the region below the Fermi surface 
$p \leq k_f$) is completed by adding to the terms in eqs.(52-55) the
contributions  from $1\pi$-exchange and iterated $1\pi$-exchange written down
in eqs.(28-36) of ref.\cite{isopot}. The imaginary isovector single-particle
$W_I(p,k_f)$ (below the Fermi surface $p \leq k_f$) has been discussed in 
section 4.2 of ref.\cite{isopot}.  

The generalization of the Hugenholtz-Van-Hove theorem \cite{vanhove} to
isospin-asymmetric nuclear matter gives a model-independent relation for the 
isovector single-particle potential $U_I(p,k_f)$ at the Fermi surface
($p=k_f)$: 
\begin{equation} U_I(k_f,k_f) = 2 A(k_f) - {k_f^2\over 3M} + {k_f^4\over 6M^3}
-{k_f\over 3} \, {\partial U(p,k_f) \over \partial p}\bigg|_{p=k_f}\,.
\end{equation}      
The second and third term on the right hand side just subtract non-interacting
(kinetic) contributions from the asymmetry energy $A(k_f)$. We find at 
saturation density $k_{f0}=261.6\,$MeV an isovector single-particle potential 
of $U_I(k_{f0},k_{f0})= 40.4\,$MeV. This is consistent with the value $U_1 
\simeq 40\,$MeV \cite{hodgson} deduced from nucleon-nucleus scattering in the
framework of the optical model. The generalized Hugenholtz-Van-Hove theorem
eq.(56) serves also as an excellent check on our analytical and numerical
calculations.
     
\begin{figure}
\begin{center}
\includegraphics[scale=0.55,clip]{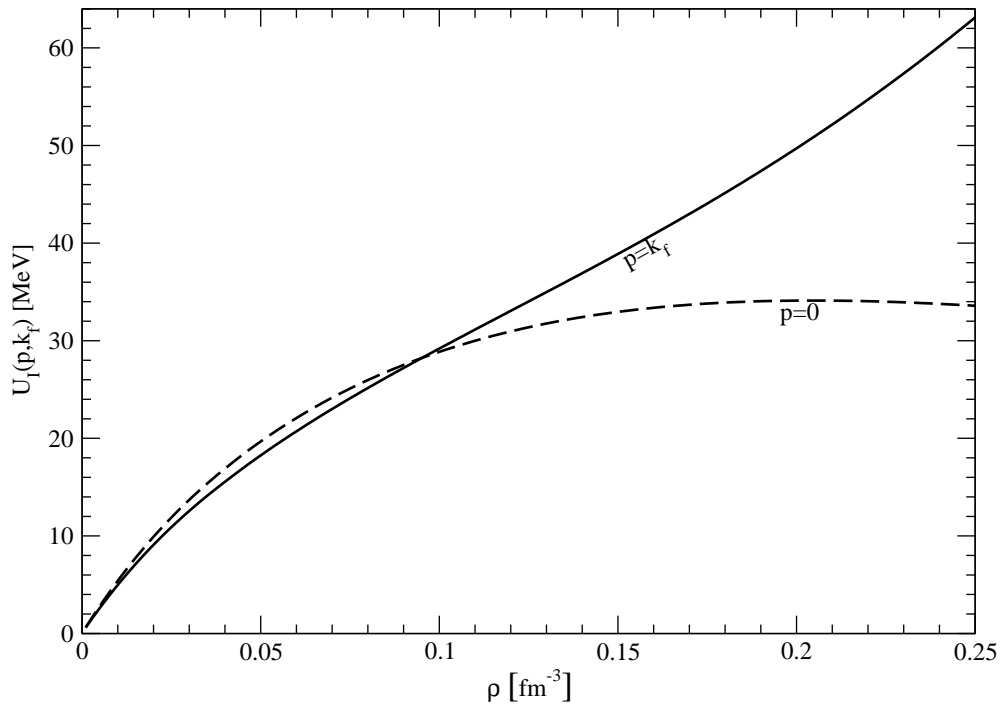}
\end{center}\vspace{-0.2cm}
\caption{The real isovector single-particle potential $U_I(p,k_f)$ as a 
function of the nucleon density $\rho= 2k_f^3/3\pi^2$. The dashed and full 
curves correspond to the sections $p=0$ (bottom of the Fermi sea) and $p=k_f$
(at the Fermi surface), respectively.} 
\end{figure}

Fig.\,12 shows the real isovector single-particle potential $U_I(p,k_f)$ as 
a function of the nucleon density $\rho= 2k_f^3/3\pi^2$. The dashed and full 
curves correspond to the sections $p=0$ (bottom of the Fermi sea) and $p=k_f$
(at the Fermi surface), respectively. One observes a splitting of both curves
which sets in at $\rho \simeq 0.10\,$fm$^{-3}$ and increases with the
density. At saturation density $\rho_0 = 0.157\,$fm$^{-3}$ the difference
between the (real) isovector single-particle potential at $p=k_{f0}$ and
$p=0$ is $U_I(k_{f0},k_{f0}) - U_I(0,k_{f0})=7.1\,$MeV. This is much less
than the analogous difference for the (real) isoscalar single-particle 
potential $U(k_{f0},k_{f0}) - U(0,k_{f0})=26.4\,$MeV (see also Fig.\,4). 

The full line in Fig.\,13 shows the (real) isovector single-particle potential 
$U_I(p,k_{f0})$ at saturation density $k_{f0}= 261.6\,$MeV as a function of 
the nucleon momentum $p$. In the region below the Fermi surface $p\leq k_{f0}$ 
the momentum dependence of this curve is weaker than that of the 
(real) isoscalar single-particle potential $U(p,k_{f0})$ shown by the left
half of the full line in Fig.\,4. Finally, we note that the generalized 
Hugenholtz-Van-Hove theorem eq.(56) (which allows for an alternative 
determination of $U_I(k_f,k_f)$ shown in Fig.\,12) holds with high 
numerical precision in our calculation.

\begin{figure}
\begin{center}
\includegraphics[scale=0.55,clip]{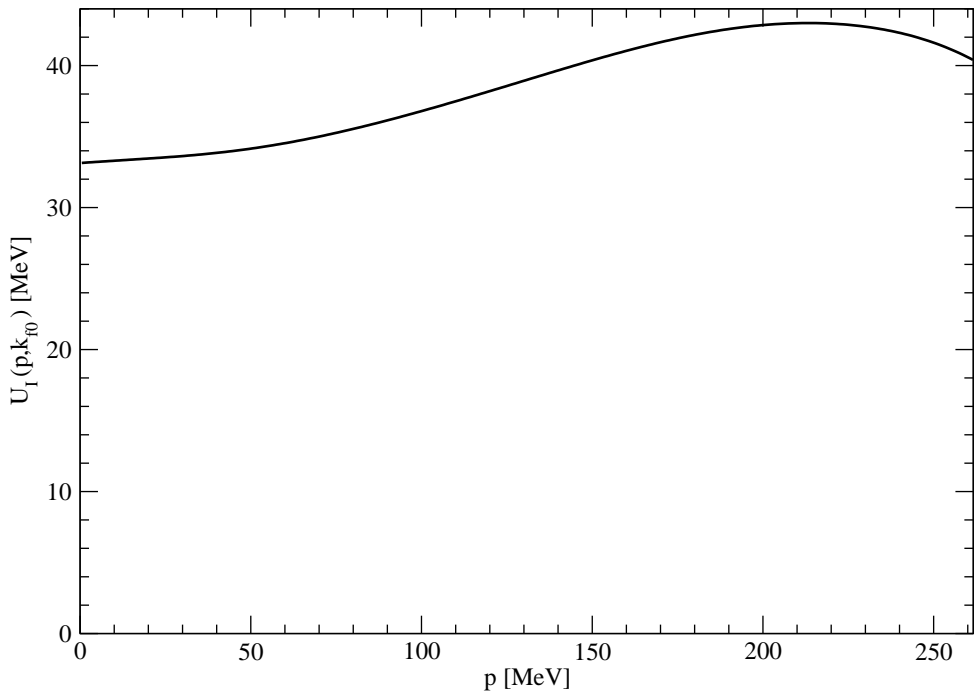}
\end{center}\vspace{-0.2cm}
\caption{The full line shows the real isovector single-particle potential 
$U_I(p,k_{f0})$ at saturation density $k_{f0}= 261.6\,$MeV as a function of 
the nucleon momentum $p$.} 
\end{figure}

\section{Summary and concluding remarks}
In this work we have extended a recent three-loop calculation of nuclear 
matter in chiral perturbation theory by including the effects from two-pion
exchange with single and double virtual $\Delta(1232)$-isobar excitation. 
In the spirit of an effective field theory, we have encoded all short-distance 
contributions (from the high-momentum parts of the pion-loops integrals etc.) 
in a few adjustable contact-coupling constants. We have investigated a wide 
variety of nuclear properties in this framework. The empirical saturation 
point of isospin-symmetric nuclear matter, $\bar E_0 = -16\,$MeV, $\rho_0 = 
0.157\,$fm$^{-3}$, can be well reproduced by adjusting the strength of a 
two-body term linear in density and weakening an emerging  three-body term 
quadratic in density. The various constraints set by 
empirical nuclear matter properties (saturation point and potential depth
lead to the (minimal) parameter choice $B_5=0$ and $\zeta= -3/4$, with little
freedom for further variation. The momentum dependence of the real 
single-particle potential $U(p,k_f)$ is improved significantly by including 
the chiral $\pi N\Delta$-dynamics. As a consequence the critical 
temperature of the liquid-gas phase transition gets lowered to the realistic 
value $T_c \simeq 15\,$MeV. The isospin properties of nuclear matter get also
substantially improved by including the chiral $\pi N\Delta$-dynamics. The 
energy per particle of pure neutron matter $\bar E_n(k_n)$ and the asymmetry 
energy $A(k_f)$ now show a monotonic growth with density. In the density
regime $\rho=2\rho_n<0.2\,$fm$^{-3}$ relevant for conventional nuclear
physics, we find good agreement with sophisticated many-body calculations and 
(semi)-empirical values. 

In passing we note that the inclusion of the chiral $\pi N \Delta$-dynamics 
guarantees the spin-stability of nuclear matter \cite{progress}. These 
improvements can be traced back to repulsive two-body Fock terms as well
as three-body terms with a very specific density and momentum dependence.     
Open questions concerning the role of yet higher orders in the small momentum 
expansion and its convergence remain and should be further explored.

Our calculation takes seriously the fact that there exist two hadronic scales, 
the pion mass $m_\pi = 135\,$MeV and the delta-nucleon mass splitting $\Delta 
= 293\,$MeV, which are smaller than or comparable to the Fermi momentum
$k_{f0}\simeq 262\,$MeV of equilibrated nuclear matter. Propagation effects of 
quasi-particles associated with these ''light'' scales are resolvable. 
Therefore pions and $\Delta$-isobars must be included as explicit degrees of 
freedom in the nuclear many-body problem. Controlled by a systematic expansion 
in small scales ($k_f,m_\pi,\Delta$), the dynamics of the interacting 
$\pi N\Delta$-system is worked out up to three-loop order. In this effective
field theory approach the basic mechanism for nuclear binding and saturation 
are attractive $2\pi$-exchange interactions of the van-der-Waals type on which
Pauli-blocking acts in the nuclear medium. Most other phenomenological 
approaches ignore these ''light'' physical degrees of freedom and parameterize
the relevant low-energy dynamics in terms of strongly coupled heavy scalar and
vector bosons ($\sigma,\,\omega,\,\rho,\,\delta$, etc.). Their propagation
takes place on length scales of $0.5\,$fm or less and can therefore not be
resolved in the domain relevant to nuclear physics. We are guided instead by a
change of paradigm, namely that the nuclear many-body problem involves the
separation of scales that is characteristic for low-energy QCD and its
(chiral) symmetry breaking pattern.

\section*{Appendix: Real single-particle potential above the Fermi surface} 
In this appendix we summarize the continuation of the real single-particle 
potential $U(p,k_f)$ calculated in section 3 of ref.\cite{pot} into the region
above the Fermi surface $p>k_f$. The expressions eqs.(8,9) in ref.\cite{pot} 
for the two-body potentials from the $1\pi$-exchange Fock diagram and the
iterated $1\pi$-exchange Hartree diagram remain valid without any
changes. Eq.(10) in ref.\cite{pot} for the two-body potential from the
iterated $1\pi$-exchange Fock diagram gets replaced by:
\begin{eqnarray} U_2(p,k_f) &=& {g_A^4Mm_\pi^4 \over (4\pi)^3 f_\pi^4} \bigg\{
u^3 + {3 \over 4x} \int_{(x-u)/2}^{(u+x)/2}\!\!d\xi\,{u^2-(2\xi-x)^2\over 1+
2\xi^2} \nonumber \\ && \times \Big[(1+8\xi^2 + 8\xi^4) \arctan\xi-(1+4\xi^2)
\arctan 2\xi \Big] \bigg\} \,, \end{eqnarray} 
with the abbreviations $u=k_f/m_\pi$ and $x=p/m_\pi$. Eq.(11) in
ref.\cite{pot} for the three-body potential from the iterated $1\pi$-exchange
Hartree diagram gets  modified to: 
\begin{eqnarray} U_3(p,k_f)&=&{6g_A^4M m_\pi^4\over (4\pi f_\pi)^4} \Bigg\{
\int_{y_{\rm min}}^1  \!\!dy\bigg\{\bigg[2uxy+(u^2-x^2y^2)\ln{u+xy\over |u-xy|}
\bigg]{\cal A}_y\bigg[{2s^2+s^4\over 2( 1+s^2)}-\ln(1+s^2)\bigg] \nonumber \\ 
&&+\int_{xy-s}^{s-xy} \!\!d\xi\,\bigg[ 2u\xi+(u^2-\xi^2)\ln{u+\xi \over u-\xi} 
\bigg]\,{(xy +\xi)^5 \over [1+ (xy +\xi)^2]^2} \bigg\}\nonumber \\ &&+ 
\int_{-1}^1 \!\!dy \int_0^u \!\!d\xi \, {\xi^2 \over x} \, \bigg[{2\sigma^2 
+\sigma^4\over 1+\sigma^2}- 2\ln(1+\sigma^2)\bigg] \ln {x+\xi y\over x-\xi
y} \, \Bigg\} \,. \end{eqnarray} 
Finally,  eq.(13) in ref.\cite{pot} for the three-body potential from the 
iterated  $1\pi$-exchange Fock diagram is replaced by:
\begin{eqnarray} U_3(p,k_f) &=&{3g_A^4M m_\pi^4 \over (4\pi f_\pi)^4} \Bigg\{
{G^2(x) \over 8x^2} +\int_0^u \!\! d\xi \, G(\xi) \bigg[ 1 +{\xi^2-x^2-1
\over 4 x \xi }\ln{1+(x+\xi)^2 \over 1+(x-\xi)^2} \bigg] \nonumber \\ && + 
\int_{y_{\rm min}}^1\!\!dy\int_{y_{\rm min}}^1\!\!dz {\theta(y^2+z^2-1) \over
4\sqrt{y^2+ z^2-1}}\,{\cal A}_y\Big[s^2-\ln(1+s^2) \Big] {\cal A}_z
\Big[\ln(1+t^2)-t^2\Big] \\ && + \int_{-1}^1 \!\!dy \int_0^u \!\!d\xi \,
{\xi^2\over x} \Big[ \ln(1+\sigma^2)-\sigma^2\Big] \bigg( \ln {x+\xi y\over
x-\xi y}+{1\over R} \ln{x R+(x^2-\xi^2-1)y \xi \over x R+(1-x^2+\xi^2)y \xi} 
\bigg) \Bigg\}\,, \nonumber \end{eqnarray}
where we have again introduced the auxiliary function:
\begin{equation} G(x) = u(1+u^2+x^2) -{1\over 4x}\big[1+(u+x)^2\big] \big[1+
(u-x)^2\big] \ln{1+(u+x)^2\over 1+(u-x)^2 } \,. \end{equation} 
For the definition of the quantities $y_{\rm min}$, $s$, $\sigma$, $t$ and
$R$ and the antisymmetrization prescription ${\cal A}_y$ we refer to section
4.

\end{document}